\documentclass[floatfix,twocolumn,showpacs,preprintnumbers,amsmath,amssymb,pra,superscriptaddress,longbibliography]{revtex4-1}
\usepackage{color}
\usepackage[usenames,dvipsnames,svgnames,table]{xcolor}
\usepackage[colorlinks=true,linkcolor=blue,urlcolor=blue,citecolor=blue]{hyperref}
\usepackage{mathtools}
\usepackage{graphicx}
\usepackage{dcolumn}
\usepackage{array}
\usepackage{lipsum}
\usepackage{bm}
\usepackage{subfigure}
\usepackage{amssymb}
\usepackage{multirow}
\usepackage{tabularx}
\usepackage{amsmath}
\usepackage{braket}
\usepackage{csquotes}
\graphicspath{{plots/}}
 \usepackage{lipsum}
\usepackage{mathrsfs}
\usepackage{MnSymbol}
\newcommand{\beq}{\begin{equation}}
\newcommand{\eeq}{\end{equation}}
\newcommand{\bea}{\begin{eqnarray}}
\newcommand{\eea}{\end{eqnarray}}

\begin{document}
%\underline{Draft version:}
\title{\emph{Ab initio} path integral Monte Carlo simulations of warm dense two-component systems without fixed nodes: structural properties}

\author{Tobias Dornheim}
\email{t.dornheim@hzdr.de}
\affiliation{Center for Advanced Systems Understanding (CASUS), D-02826 G\"orlitz, Germany}
\affiliation{Helmholtz-Zentrum Dresden-Rossendorf (HZDR), D-01328 Dresden, Germany}

\author{Sebastian Schwalbe}
\affiliation{Center for Advanced Systems Understanding (CASUS), D-02826 G\"orlitz, Germany}
\affiliation{Helmholtz-Zentrum Dresden-Rossendorf (HZDR), D-01328 Dresden, Germany}

\author{Maximilian~P.~B\"ohme}
\affiliation{Center for Advanced Systems Understanding (CASUS), D-02826 G\"orlitz, Germany}
\affiliation{Helmholtz-Zentrum Dresden-Rossendorf (HZDR), D-01328 Dresden, Germany}
\affiliation{Technische  Universit\"at  Dresden,  D-01062  Dresden,  Germany}

\author{Zhandos A.~Moldabekov}
\affiliation{Center for Advanced Systems Understanding (CASUS), D-02826 G\"orlitz, Germany}
\affiliation{Helmholtz-Zentrum Dresden-Rossendorf (HZDR), D-01328 Dresden, Germany}

%\author{Thomas Gawne}
%\affiliation{Center for Advanced Systems Understanding (CASUS), D-02826 G\"orlitz, Germany}
%\affiliation{Helmholtz-Zentrum Dresden-Rossendorf (HZDR), D-01328 Dresden, Germany}

\author{Jan Vorberger}
\affiliation{Helmholtz-Zentrum Dresden-Rossendorf (HZDR), D-01328 Dresden, Germany}

%\author{Tilo~D\"oppner}
%\affiliation{Lawrence Livermore National Laboratory (LLNL), California 94550 Livermore, USA}

%\author{Frank R.~Graziani}
%\affiliation{Lawrence Livermore National Laboratory (LLNL), California 94550 Livermore, USA}

%\author{Dominik~Kraus}
%\affiliation{Institut f\"ur Physik, Universit\"at Rostock, D-18057 Rostock, Germany}
%\affiliation{Helmholtz-Zentrum Dresden-Rossendorf (HZDR), D-01328 Dresden, Germany}

\author{Panagiotis Tolias}
\affiliation{Space and Plasma Physics, Royal Institute of Technology (KTH), Stockholm, SE-100 44, Sweden}

\begin{abstract}
We present extensive new \emph{ab initio} path integral Monte Carlo (PIMC) results for a variety of structural properties of warm dense hydrogen and beryllium. To deal with the fermion sign problem---an exponential computational bottleneck due to the antisymmetry of the electronic thermal density matrix---we employ the recently proposed [\textit{J.~Chem.~Phys.}~\textbf{157}, 094112 (2022); \textbf{159}, 164113 (2023)] $\xi$-extrapolation method and find excellent agreement with exact direct PIMC reference data where available. This opens up the intriguing possibility to study a gamut of properties of light elements and potentially material mixtures over a substantial part of the warm dense matter regime, with direct relevance for astrophysics, material science, and inertial confinement fusion research.
%More formal / technical analysis of the capabilities of the $\xi$-extrapolation applied to two-component systems. \textcolor{red}{Using $N=32$, $r_s=3.23$, $r_s=2$, $\Theta=1$ as a basis.} \textcolor{blue}{Next task: analyse $N=14$ for both cases, start calcs for $r_s=3.23$}
%\textcolor{red}{Energy, pressure, EOS, momentum distribution etc deserves dedicated future publication!}
\end{abstract}
\maketitle

\section{Introduction\label{sec:introduction}}

Warm dense matter (WDM) is an extreme state that is characterized by the simultaneous presence of high temperatures, pressures, and densities~\cite{wdm_book}. In nature, WDM occurs in a host of astrophysical objects, including giant planet interiors~\cite{Benuzzi_Mounaix_2014}, brown dwarfs~\cite{becker}, and the outer layer of neutron stars~\cite{Haensel}. In addition, WDM plays an important role for technological applications such as hot-electron chemistry~\cite{Brongersma2015}, and the discovery and synthesis of exotic materials~\cite{Lazicki2021,Kraus2016,Kraus2017}. A particularly important example is given by inertial confinement fusion (ICF)~\cite{Betti2016}, where the fuel capsule (typically a mixture of the hydrogen isotopes deuterium and tritium) has to traverse the WDM regime in a controlled way to reach ignition~\cite{hu_ICF}. In fact, the recently reported~\cite{NIF_PRL_2024} net energy gain from an ICF experiment has opened up the intriguing possibility to further develop ICF into an economically viable and sustainable option for the production of clean energy~\cite{roadmap}.

Yet, the rigorous theoretical description of WDM constitutes a notoriously difficult challenge~\cite{wdm_book,review,Dornheim_review,new_POP}, as it must capture the complex interplay of Coulomb coupling and nonideality, quantum degeneracy and diffraction, as well as strong thermal excitations out of the ground state. From a theoretical perspective, these conditions are conveniently characterized in terms of the Wigner-Seitz radius (also known as \emph{density parameter} or \emph{quantum coupling parameter}) $r_s=(3/4\pi n)^{1/3}$, where $n=N/\Omega$ is the electronic number density, and the degeneracy temperature $\Theta=k_\textnormal{B}T/E_\textnormal{F}$, with $E_\textnormal{F}$ the usual Fermi energy~\cite{quantum_theory}. In the WDM regime, one has $r_s\sim\Theta\sim1$~\cite{Ott2018}, which means that there are no small parameters for an expansion~\cite{wdm_book}. 

In this situation, thermal density functional theory (DFT)~\cite{Mermin_DFT_1965} has emerged as the de-facto work horse of WDM theory as it combines a generally manageable computation cost with an often acceptable level of accuracy. At the same time, it is important to note that DFT crucially relies on external input i) to provide the required exchange--correlation (XC) functional, ii) as a benchmark for its inherent approximations. At ambient conditions where the electrons are in their ground state, the availability of highly accurate \emph{ab initio} quantum Monte Carlo (QMC) results for the uniform electron gas (UEG)~\cite{Ceperley_Alder_PRL_1980,Spink_PRB_2013,moroni,moroni2} was of paramount importance to facilitate the arguably unrivaled success of DFT with respect to the description of real materials~\cite{Jones_RMP_2015}. Yet, it is easy to see that thermal DFT simulations of WDM require as an input a thermal XC-functional that depends on both density and the temperature~\cite{karasiev_importance,computation4020016,wdm_book}, the construction of which, in turn, must be based on thermal QMC simulations at WDM conditions \cite{Moldabekov_non_empirical_hybrid}.

In this context, the \emph{ab initio} path integral Monte Carlo (PIMC) method~\cite{cep,Takahashi_Imada_PIMC_1984,Berne_JCP_1982} constitutes a key concept. On the one hand, PIMC is, in principle, capable of providing an exact solution to a given quantum many-body problem of interest for arbitrary temperatures; this has given insights into important phenomena such as superfluidity~\cite{Pollock_PRB_1987,cep,Dornheim_PRB_2015} and Bose-Einstein condensation~\cite{PhysRevA.70.053614,doi:10.7566/JPSJ.85.053001}.
On the other hand, PIMC simulations of Fermi systems such as the electrons in WDM are afflicted with the notorious fermion sign problem (FSP)~\cite{dornheim_sign_problem,Dornheim_grand_2021,troyer}. It leads to an exponential increase in the required compute time with increasing the system size $N$ or decreasing the temperature $T$, and has prevented PIMC simulations over a substantial part of the WDM regime.

This unsatisfactory situation has sparked a surge of developments in the field of fermionic QMC simulations targeting WDM~\cite{Brown_PRL_2013,Schoof_PRL_2015,Dornheim_NJP_2015,dornheim_prl,Malone_JCP_2015,Malone_PRL_2016,Joonho_JCP_2021,Yilmaz_JCP_2020,Dornheim_CPP_2019}, see Refs.~\cite{Dornheim_POP_2017,review,Dornheim_review} and references therein. These efforts have culminated in the first accurate parametrizations of the XC-free energy of the warm dense UEG~\cite{ksdt,groth_prl,review,Karasiev_status_2019,Roepke_PRE_2024}, which can be used as input for thermal DFT simulations of WDM~\cite{Sjostrom_PRB_2014,karasiev_importance,kushal,Moldabekov_JCTC_BoundState_2023,Moldabekov_JCTC_2023, moldabekov2021relevance, Moldabekov_PRB2022} on the level of the local density approximation. While being an important milestone, the development of highly accurate PIMC simulations of real WDM systems where the positively charged nuclei are not approximated by a homogeneous background remains of the utmost importance for many reasons, e.g.: i) it is important to benchmark different XC-functionals for thermal DFT simulations in realistic situations; ii) PIMC results can be used to construct more sophisticated thermal XC-functionals, e.g.~via the fluctuation--dissipation theorem~\cite{pribram}; iii) PIMC gives one straightforward access to the imaginary-time density--density correlation function (ITCF) $F_{ee}(\mathbf{q},\tau)$, which is of key importance for the interpretation of X-ray Thomson scattering (XRTS) experiments with WDM~\cite{Dornheim_T_2022,Dornheim_T2_2022,Dornheim_insight_2022,Dornheim_review,boehme2023evidence,dornheim2023xray}; iv) while DFT is an effective single-electron theory, PIMC allows one to compute many-body correlation functions and related nonlinear effects~\cite{Dornheim_JPSJ_2021,Dornheim_review,Dornheim_PRL_2020,Dornheim_PRR_2021,Dornheim_JCP_ITCF_2021,Tolias_EPL_2023}.

As a consequence, a number of strategies to attenuate the FSP in PIMC simulations have been discussed in the literature~\cite{Chin_PRE_2015,Dornheim_NJP_2015,Dornheim_JCP_2015,Dornheim_CPP_2019,review,Hirshberg_JCP_2020,Dornheim_JCP_2020,Ceperley1991,Brown_PRL_2013,Driver_PRL_2012,Militzer_PRL_2015,Xiong_JCP_2022,Xiong_PRE_2023,Dornheim_JCP_2023,Dornheim_JPCL_2024}. Note that complementary QMC methods like configuration PIMC~\cite{Schoof_CPP_2011,Schoof_CPP_2015,Schoof_PRL_2015,Groth_PRB_2016,review,Yilmaz_JCP_2020} and density matrix QMC~\cite{Malone_JCP_2015,Malone_PRL_2016,Blunt_PRB_2014} lie beyond the scope of the present work and have been covered elsewhere. Three decades ago, Ceperley~\cite{Ceperley1991} suggested a reformulation of the PIMC method, where the sampling of any negative contributions can be avoided by prohibiting paths that cross the nodal surfaces of the thermal density matrix. This \emph{fixed-node approximation}, also known as restricted PIMC (RPIMC) in the literature, is both formally exact and sign-problem free. Unfortunately, the exact nodal surface of an interacting many-fermion system is generally a-priori unknown, and one has to rely on de-facto uncontrolled approximations in practice. For the warm dense UEG, Schoof \emph{et al.}~\cite{Schoof_PRL_2015} have reported systematic errors of $\sim10\%$ in the XC-energy at high density ($r_s=1$), whereas Dornheim \emph{et al.}~\cite{Dornheim_PRB_nk_2021} have found better agreement for the momentum distribution for $r_s\geq 2$.
In addition, the nodal restrictions on which RPIMC is based break the usual imaginary-time translation invariance and, thus, prevent the straightforward estimation of imaginary-time correlation functions. In spite of these shortcomings, the RPIMC method has been successfully applied by Militzer and coworkers~\cite{Militzer_2008,Militzer_PRE_2001,Militzer_PRL_2015,Driver_PRL_2012} to a variety of WDM systems, and their results form the basis for an extensive equation-of-state database~\cite{Militzer_PRE_2021}.

A different line of thought has been pursued by Takahashi and Imada~\cite{Takahashi_Imada_PIMC_1984}, who have suggested to use inherently anti-symmetric imaginary-time propagators, i.e., determinants. This leads to the blocking (grouping together) of positive and negative contributions to the partition function into a single term, thereby alleviating the sign problem. Similar considerations have been used by Filinov \emph{et al.} for PIMC simulations of the UEG~\cite{Filinov_PRE_2015}, warm dense hydrogen~\cite{Filinov_CPP_2004}, and exotic quark-gluon plasmas~\cite{FILINOV20121096}. In practice, this strategy works well if the number of imaginary-time propagators $P$ is small enough so that the thermal wavelength of a single imaginary-time slice $\lambda_\epsilon=\sqrt{2\pi\epsilon}$, with $\epsilon=\beta/P$, is comparable to the average interparticle distance $\overline{r}$. The key point is thus to combine the determinant with a high-order factorization of the density matrix~\cite{sakkos_JCP_2009}, which allows for sufficient accuracy even for very small $P$. This is the basic idea of the permutation blocking PIMC (PB-PIMC) method~\cite{Dornheim_NJP_2015,Dornheim_JCP_2015,Dornheim_PRB_2016,review,dornheim_pre,Dornheim_CPP_2019}, which has very recently been applied to the simulation of warm dense hydrogen by Filinov and Bonitz~\cite{Filinov_PRE_2023}. While being arguably less uncontrolled than the RPIMC method, the PB-PIMC idea has three main bottlenecks: i) even though the FSP is significantly attenuated, the method still scales exponentially with $\beta$ and $N$; ii) the convergence with $P$ is often difficult to ensure, especially at low $T$ where this would be most important; iii) it is difficult to resolve imaginary-time properties due to the necessarily small number of imaginary-time slices.

A third route has been recently suggested by Xiong and Xiong~\cite{Xiong_JCP_2022,Xiong_PRE_2023} based on the path integral molecular dynamics simulation of fictitious identical particles that are defined by the continuous spin-variable $\xi\in[-1,1]$ [cf.~Eq.~(\ref{eq:Z}) below]. To be more specific, they have proposed to carry out simulations in the sign-problem free domain of $\xi\geq0$, and to subsequently extrapolate to the fermionic limit of $\xi=-1$. This $\xi$-extrapolation method has subsequently been adapted to PIMC simulations of the warm dense UEG by Dornheim \emph{et al.}~\cite{Dornheim_JCP_2023,Dornheim_JPCL_2024}, who have found that it works remarkably well for weak to moderate levels of quantum degeneracy. Although such an extrapolation is empirical, it combines a number of strong advantages in practice. First and foremost, only effects due to quantum statistics have to be extrapolated. These are known to be local in the case of fermions~\cite{Kohn_PNAS_2005}, which means that it is possible to verify the applicability (or lack thereof) of the $\xi$-extrapolation method for relatively small systems (e.g.~$N\sim4,...,14$) before applying it to larger numbers of particles where no direct check is possible~\cite{Dornheim_JCP_2023}. Second, the method has no sign-problem, which allows one to simulate very large numbers of electrons~\cite{Dornheim_JPCL_2024}. Finally, it gives one access to all observables that can be computed with direct PIMC, including the ITCF.
This has very recently allowed Dornheim \emph{et al.}~\cite{Dornheim_Science_2024} to compare extensive PIMC simulations of warm dense Be to XRTS experiments carried out at the National Ignition Facility (NIF)~\cite{Tilo_Nature_2023}, resulting in an unprecedented consistency between theory and experiment without the need for any empirical parameters.

In the present work, we report a detailed study of the application of the $\xi$-extrapolation method to warm dense two-component plasmas, focusing on various structural characteristics of hydrogen and beryllium. The paper is organized as follows. In Sec.~\ref{sec:theory}, we introduce the relevant theoretical background, including a definition of the all-electron Hamiltonian governing the entire two-component system (\ref{sec:Hamiltonan}), a brief introduction to the \emph{ab initio} PIMC method (\ref{sec:PIMC}) and a subsequent overview of the $\xi$-extrapolation method (\ref{sec:xi_extrapolation}). In Sec.~\ref{sec:results}, we present our extensive new simulation results, starting with the fermion sign problem (\ref{sec:FSP_results}); interestingly, it is substantially more severe for two-component plasmas than for the UEG at the same conditions, reflecting the more complex physics in a real WDM system. In Secs.~\ref{sec:H_metallic} and \ref{sec:H_solid}, we present a detailed analysis of hydrogen at the electronic Fermi temperature ($\Theta=1$) at a metallic ($r_s=2$, $\rho=0.34\,$g/cc, $T=12.53\,$eV) and solid ($r_s=3.23$, $\rho=0.08\,$g/cc, $T=4.80\,$eV) density. In Sec.~\ref{sec:Be_results}, we consider the substantially more complex case of strongly compressed ($r_s=0.93$, $\rho=7.49\,$g/cc) Be at $T=100\,$eV ($\Theta=1.73$), which is relevant e.g.~for experiments at the NIF~\cite{MacDonald_POP_2023,Tilo_Nature_2023,Dornheim_Science_2024}. Remarkably, the $\xi$-extrapolation method is even capable of reproducing the correct interplay of XC-effects with double occupation of the atomic K-shell as they manifest in observables such as the spin-resolved pair correlation function. Finally, we consider the spatially resolved electronic density in the external potential of a fixed ion snapshot in Sec.~\ref{sec:Be_SNAP_results}. The paper is concluded by a summary and outlook in Sec.~\ref{sec:summary}.

\section{Theory\label{sec:theory}}

\subsection{Hamiltonian\label{sec:Hamiltonan}}

Throughout this work, we restrict ourselves to the fully unpolarized case where $N_\uparrow = N_\downarrow = N/2$ with $N$ being the total number of electrons in the system. Moreover, we consider effectively charge neutral systems where $N=ZN_\textnormal{atom}$ with $Z$ being the nuclear charge and $N_\textnormal{atom}$ the total number of nuclei.
The corresponding Hamiltonian governing the behaviour of the thus defined two-component plasma then reads
\begin{widetext}
\begin{eqnarray}\label{eq:H}
    \hat{H} = -\frac{1}{2} \sum_{l=1}^N \nabla_l^2  - \frac{1}{2m_n} \sum_{l=1}^{N_\textnormal{atom}} \nabla_l^2 + \sum_{l<k}^N W_\textnormal{E}(\hat{\mathbf{r}}_l,\hat{\mathbf{r}}_k) + Z^2 \sum_{l<k}^{N_\textnormal{atom}} W_\textnormal{E}(\hat{\mathbf{I}}_l,\hat{\mathbf{I}}_k) - Z \sum_{k=1}^N\sum_{l=1}^{N_\textnormal{atom}} W_\textnormal{E}(\hat{\mathbf{I}}_l,\hat{\mathbf{r}}_k)\ ,
\end{eqnarray}
\end{widetext}
where $\hat{\mathbf{r}}$ and $\hat{\mathbf{I}}$ denote electron and nucleus position operators,
and we assume Hartree atomic units (i.e., $m_e=1$) throughout this work. The pair interaction is given by the usual Ewald potential, where we follow the convention given by Fraser \emph{et al.}~\cite{Fraser_PRB_1996}.

\subsection{Path integral Monte Carlo\label{sec:PIMC}}

The \emph{ab initio} PIMC method constitutes one of the most successful tools in statistical physics and quantum chemistry.
Since more detailed introductions to PIMC~\cite{cep,Takahashi_Imada_PIMC_1984,Berne_JCP_1982} and its algorithmic implementation~\cite{boninsegni1,Dornheim_PRB_nk_2021} have been presented in the literature, we restrict ourselves to a brief summary of the main ideas. As a starting  point, we consider the canonical (i.e., number of particles $N$, volume $\Omega$, and inverse temperature $\beta=1/k_\textnormal{B}T$ are fixed) partition function in coordinate representation,
\begin{widetext}
\begin{eqnarray}\label{eq:Z}
    Z_{N,\Omega,\beta} = \frac{1}{N_\uparrow!N_\downarrow!} \sum_{\sigma_{N_\uparrow}\in S_{N_\uparrow}}\sum_{\sigma_{N_\downarrow}\in S_{N_\downarrow}} \xi^{N_{pp}}  \int\textnormal{d}\mathbf{R} \bra{\mathbf{R}} e^{-\beta\hat{H}} \ket{\hat{\pi}_{\sigma_{N_\uparrow}}\hat{\pi}_{\sigma_{N_\downarrow}}\mathbf{R}}
\end{eqnarray}
\end{widetext}
where $\mathbf{R}$ contains the coordinates of all electrons and nuclei. The double sum over all possible permutations of coordinates of the spin-up and spin-down electrons is required to properly realize the fermionic antisymmetry of the thermal density matrix, where the exponent $N_{pp}$ corresponds to the number of pair permutations that is required to realize a particular combination of permutation elements $\sigma_{N_\uparrow}$ and $\sigma_{N_\downarrow}$.
In general, the cases $\xi=1$, $\xi=0$, and $\xi=-1$ correspond to Bose-, Boltzmann-, and Fermi-statistics, with the latter applying to the electrons studied in the present work. We note that we treat the nuclei as distinguishable \emph{Boltzmann particles} (i.e., \emph{boltzmannons}), which is appropriate for the conditions studied in the present work. Further note that this does still take into account nuclear quantum effects due to the finite extension of the nucleonic paths, which are however small ($\sim0.1\%$) in the warm dense matter regime even for hydrogen.

The problem with Eq.~(\ref{eq:Z}) concerns the evaluation of the matrix elements of the density operator $\hat\rho = e^{-\beta\hat{H}}$ that is not directly possible in practice, because the potential and kinetic contributions to the full Hamiltonian $\hat{H}=\hat{V} + \hat{K}$ do not commute. The usual workaround is to evoke the exact semi-group property of the density operator combined with a Trotter decomposition~\cite{Trotter} of the latter, leading to the evaluation of $P$ density matrices at a $P$-times higher temperature. In the case of electron--ion systems, an additional obstacle is given by the diverging Coulomb attraction between electrons and ions at short distances; this prevents a straightforward application of the Trotter formula, which only holds for potentials that are bounded from below.
We overcome this problem by using the well-known \emph{pair approximation}~\cite{MILITZER201688,POLLOCK198849,Bohme_PRE_2023,cep} that is based on the exact solution of the isolated electron--ion two-body problem, and utilize the implementation presented in Ref.~\cite{Bohme_PRE_2023}. In essence, the quantum many-body problem defined by Eq.~(\ref{eq:Z}) has then been mapped onto an ensemble of quasi-classical ring polymers with $P$ segments~\cite{Chandler_JCP_1981}; these are the eponymous paths of the PIMC method. Alternatively, one can say that each quantum particle is now represented by an entire path of length $\tau=\beta$ along the imaginary time $t=-i\hbar\tau$ with $P$ discrete imaginary-time steps. Therefore, PIMC gives one straightforward access to a number of imaginary-time correlation functions~\cite{Dornheim_JCP_ITCF_2021,Rabani_PNAS_2002,boninsegni1,Filinov_PRA_2012}, with the density--density correlator $F_{ab}(\mathbf{q},\tau)=\braket{\hat{n}_a(\mathbf{q},\tau)\hat{n}_b(-\mathbf{q},0)}$ between species $a$ and $b$ being a particularly relevant example for WDM research~\cite{dornheim_dynamic,dynamic_folgepaper,Hamann_PRB_2020,Dornheim_T_2022,Dornheim_T2_2022,Dornheim_insight_2022}. The PIMC formulation becomes exact in the limit of $P\to\infty$, and the convergence with $P$ has been carefully checked. We find that $P=200$ proves sufficient for all cases studied here.

The basic idea of the PIMC method is to randomly generate all possible path configurations $\mathbf{X}$ (where the meta variable $\mathbf{X}$ contains both the coordinates of electrons and nuclei) using an implementation of the celebrated Metropolis algorithm~\cite{metropolis}. We note that this also requires the sampling of different permutation topologies, which can be accomplished efficiently via the worm algorithm that was introduced by Boninsegni \emph{et al.}~\cite{boninsegni1,boninsegni2}. In practice, we use the extended ensemble algorithm from Ref.~\cite{Dornheim_PRB_nk_2021} which is implemented in the \texttt{ISHTAR} code~\cite{ISHTAR}.

\subsection{The $\xi$-extrapolation method\label{sec:xi_extrapolation}}

In the case of fermions, the factor of $(-1)^{N_{pp}}$ leads to a cancellation of positive and negative terms, which is the root cause of the notorious fermion sign problem~\cite{dornheim_sign_problem,Dornheim_grand_2021,troyer}; it results in an exponential increase in the compute time with $N$ and $\beta$. In other words, the signal-to-noise ratio of the direct PIMC method vanishes for large systems, and at low temperatures. As a partial remedy of this bottleneck, Xiong and Xiong~\cite{Xiong_JCP_2022} have suggested to carry out path integral molecular dynamics simulations of fictitious identical particles where $\xi\in[-1,1]$ is treated as a continuous variable. It is straightforward to carry out highly accurate simulations in the sign-problem free domain of $\xi\in[0,1]$; one can then extrapolate to the fermionic limit of $\xi=-1$ using the empirical relation
\begin{eqnarray}\label{eq:fit}
    A(\xi) = a_0 + a_1\xi + a_2\xi^2\ .
\end{eqnarray}
For completeness, we note that an alternative extrapolation method has been introduced very recently~\cite{Xiong_PRE_2023}, but the possibility of extending it to observables beyond the total energy remains a subject for dedicated future works.

\begin{figure}\centering
\includegraphics[width=0.45\textwidth]{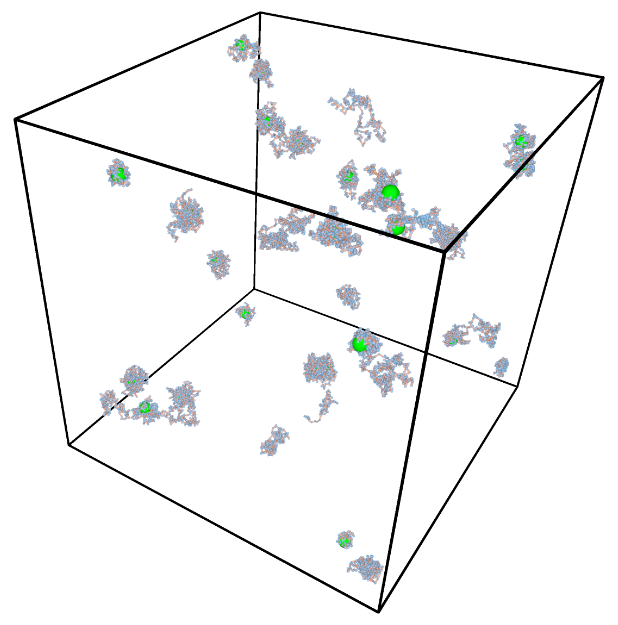}
\caption{\label{fig:BeFollowUp} Snapshot of an all-electron PIMC simulation of $N_\textnormal{atom}=25$ Be atoms at $r_s=0.93$ ($\rho=7.49\,$g/cc) and $\Theta=1.73$ ($T=100\,$eV). The green orbs depict the approximately point-like nuclei, and the red-blue paths represent the quantum degenerate (i.e., smeared out) electrons.
}
\end{figure} 

Subsequently, Dornheim \emph{et al.}~\cite{Dornheim_JCP_2023} have adapted the $\xi$-extrapolation approach for PIMC simulations of the warm dense UEG, and found that is works remarkably well for weak to moderate quantum degeneracy. 
A particular practical advantage of this method is that it can be rigorously verified for small systems [$N=\mathcal{O}(1)-\mathcal{O}(10)$], where comparison with exact direct PIMC calculations is feasible. Then, a breakdown of the $\xi$-extrapolation [i.e., Eq.~(\ref{eq:fit})] for larger systems at the same conditions is rendered highly unlikely due to the well-known local nature of quantum statistics effects for fermions~\cite{Kohn_PNAS_2005}. Other effects, such as the interplay of long-range Coulomb coupling with the quantum delocalization of the electrons is still fully taken into account by the PIMC simulations in the sign-problem free domain of $\xi\in[0,1]$ for larger systems.
This is illustrated in Fig.~\ref{fig:BeFollowUp}, where we show a snapshot of a PIMC simulation of $N_\textnormal{atom}=25$ Be atoms at $r_s=0.93$ ($\rho=7.49\,$g/cc) and $\Theta=1.73$ ($T=100\,$eV). The green orbs represent the Be nuclei, which are indistinguishable from classical point particles at these conditions, even though this assumption is not hardwired into our set-up, see Sec.~\ref{sec:PIMC} above.
The red-blue paths represent the quantum degenerate electrons, with their extension being proportional to the thermal wavelength $\lambda_\beta=\sqrt{2\pi\beta}$. Thus, effects such as quantum diffraction are covered on all length scales, which has recently allowed four of us to present unprecedented simulations of the warm dense UEG with up to $N=1000$ electrons~\cite{Dornheim_JPCL_2024}.
Furthermore, the $\xi$-extrapolation method has been used in Ref.~\cite{Dornheim_Science_2024} to carry out extensive PIMC simulations of strongly compressed Be ($\rho=7.5-30\,$g/cc), resulting in excellent agreement with XRTS measurements taken at the NIF~\cite{Tilo_Nature_2023}. Here, we present a much more detailed technical analysis of this approach, and apply it to  hydrogen and beryllium for a set of different conditions.

\section{Results\label{sec:results}}

All PIMC results that are presented in this work are freely available in an online repository, see Ref.~\cite{repo}.

\subsection{Fermion sign problem\label{sec:FSP_results}}

\begin{figure}\centering
\includegraphics[width=0.45\textwidth]{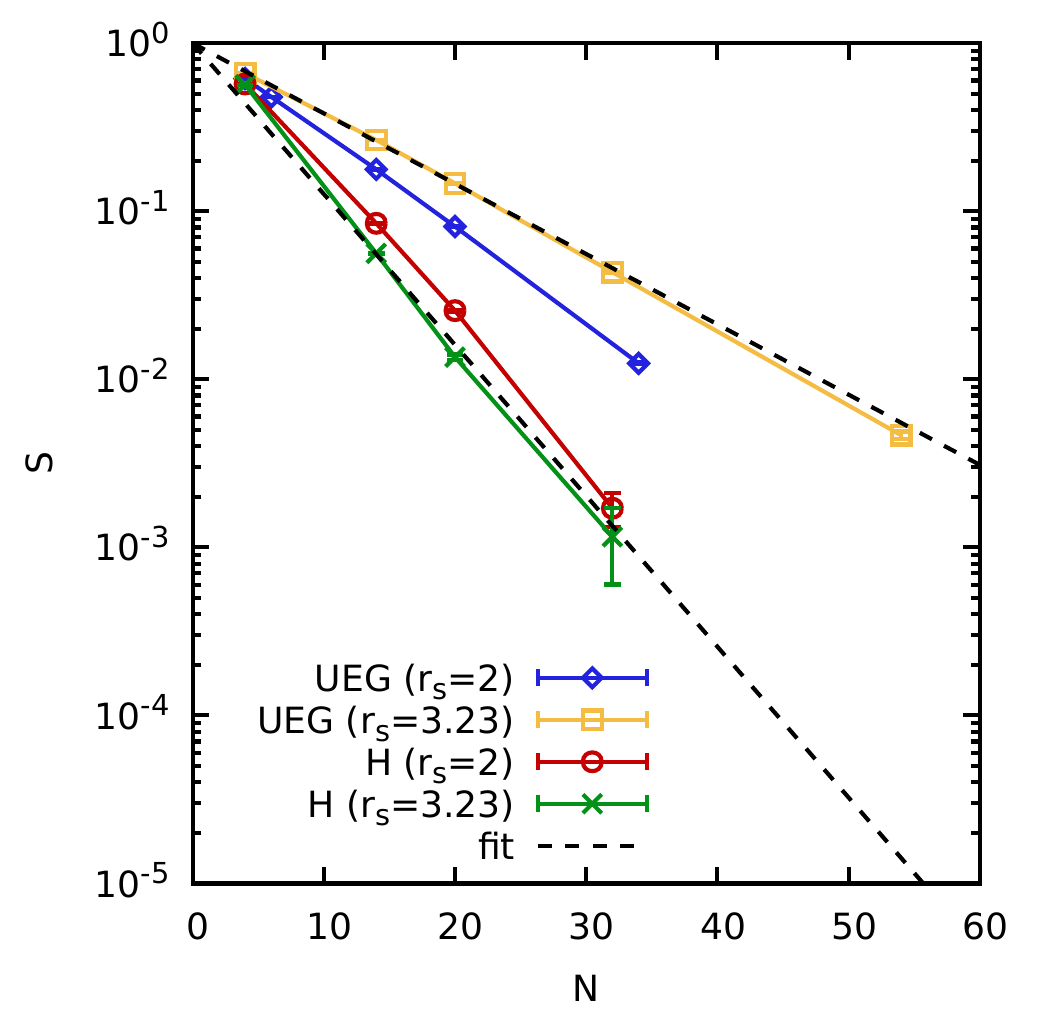}
\caption{\label{fig:Hydrogen_Sign_theta1} PIMC results for the dependence of the average sign $S$ on the system size $N$ at the electronic Fermi temperature, $\Theta=1$. Blue diamonds (green crosses): UEG at $r_s=2$ ($r_s=3.23$); red circles (black stars): hydrogen at $r_s=2$ ($r_s=3.23$). Dashed black: exponential fit to the data at $r_s=3.23$. The UEG data for $r_s=2$ are partially taken from Ref.~\cite{Dornheim_JCP_Force_2022}.
}
\end{figure} 

The FSP constitutes the main bottleneck for PIMC simulations of WDM systems. It is a direct consequence of the cancellation of positive and negative contributions to the partition function Eq.~(\ref{eq:Z}) due to the antisymmetry of the thermal density matrix under the exchange of particles. This cancellation is conveniently characterized by the \emph{average sign} $S$ that is inversely proportional to the relative 
uncertainty of a given observable~\cite{dornheim_sign_problem}. 
In Fig.~\ref{fig:Hydrogen_Sign_theta1}, we show the dependence of $S$ on the number of electrons $N$ based on PIMC calculations. The blue diamonds and yellow squares have been obtained for the UEG at $\Theta=1$ and $r_s=2$ and $r_s=3.23$, respectively. For an ideal Fermi gas, the degree of quantum degeneracy would be exclusively a function of $\Theta$ and the results independent of the density. For the UEG, $r_s$ serves as a quantum coupling parameter~\cite{quantum_theory}. The stronger coupling at the solid density thus leads to an effective separation of the electrons within the PIMC simulation, leading to a decreased probability for the formation of permutation cycles~\cite{Dornheim_permutation_cycles}. In other words, a sparser electron gas is less quantum degenerate than a dense electron gas, resulting in the well-known monotonic increase of $S$ with $r_s$ for the UEG. In addition, we find the expected exponential decrease of the sign with the number of electrons that is well reproduced by the exponential fit to the yellow squares, and which is ultimately responsible for the exponentially vanishing signal-to-noise ratio for direct PIMC simulations of fermions. In practice, simulations are generally feasible for $S\sim\mathcal{O}\left(10^{-1}\right)$.

Let us next consider the red circles, which show PIMC results for hydrogen at $r_s=2$ (and $\Theta=1$). Evidently, the average sign is systematically lower compared to the UEG at the same conditions. This is a direct consequence of the presence of the protons, leading to local inhomogeneities. To be more specific, electrons tend, on average, to cluster around the protons; this even holds at conditions where the majority of electrons can be considered as effectively \emph{unbound}. The reduced average distance between the electrons then leads to an increased sampling of permutation cycles and, therefore, a reduced average sign compared to the UEG. 

Finally, the green crosses show results for hydrogen at $r_s=3.23$. Interestingly, we find the opposite trend compared to the UEG: the average sign decreases with increasing $r_s$. In other words, quantum degeneracy effects are even somewhat more important at lower density compared to the high-density case. This is a consequence of the more complex physics in real two-component plasmas compared to the UEG, leading to two competing trends for hydrogen. On the one hand, increasing the density compresses the electronic component, making exchange effects more important. On the other hand, reducing the density leads to the formation of $H_2$ molecules, where the two electrons are in very close proximity. In fact, the PIMC sampling by itself cannot directly distinguish between a spin-polarized or spin-unpolarized $H_2$ molecule. As we will see below, the correct predominance of the unpolarized case is exclusively a consequence of the fermionic antisymmetry that is realized by the cancellation of positive and negative contributions in PIMC, i.e., a subsequent re-weighting of the actually sampled set of configurations. In practice, this crucial effect is nicely captured by the $\xi$-extrapolation method, making it reliable beyond the comparably simple UEG model system.

\subsection{Hydrogen: metallic density\label{sec:H_metallic}}

\begin{figure*}\centering
\includegraphics[width=0.45\textwidth]{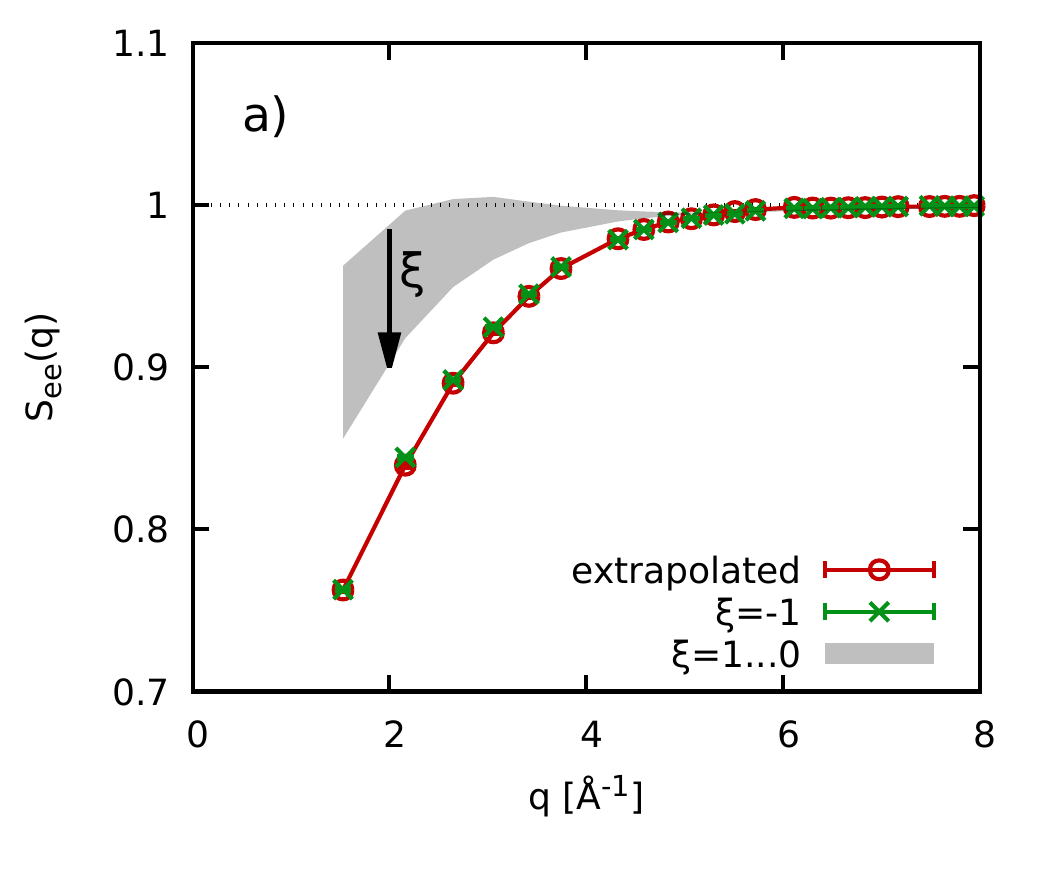}\includegraphics[width=0.45\textwidth]{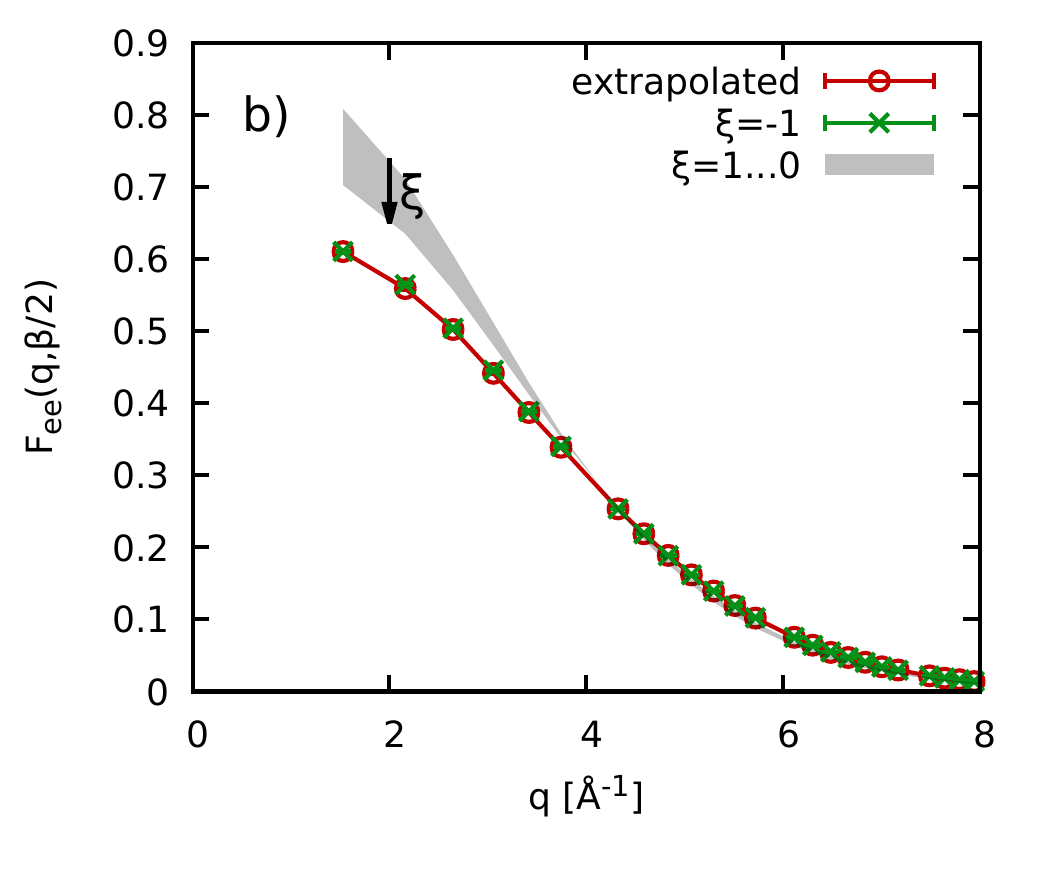}\\\vspace*{-1.25cm}
\includegraphics[width=0.45\textwidth]{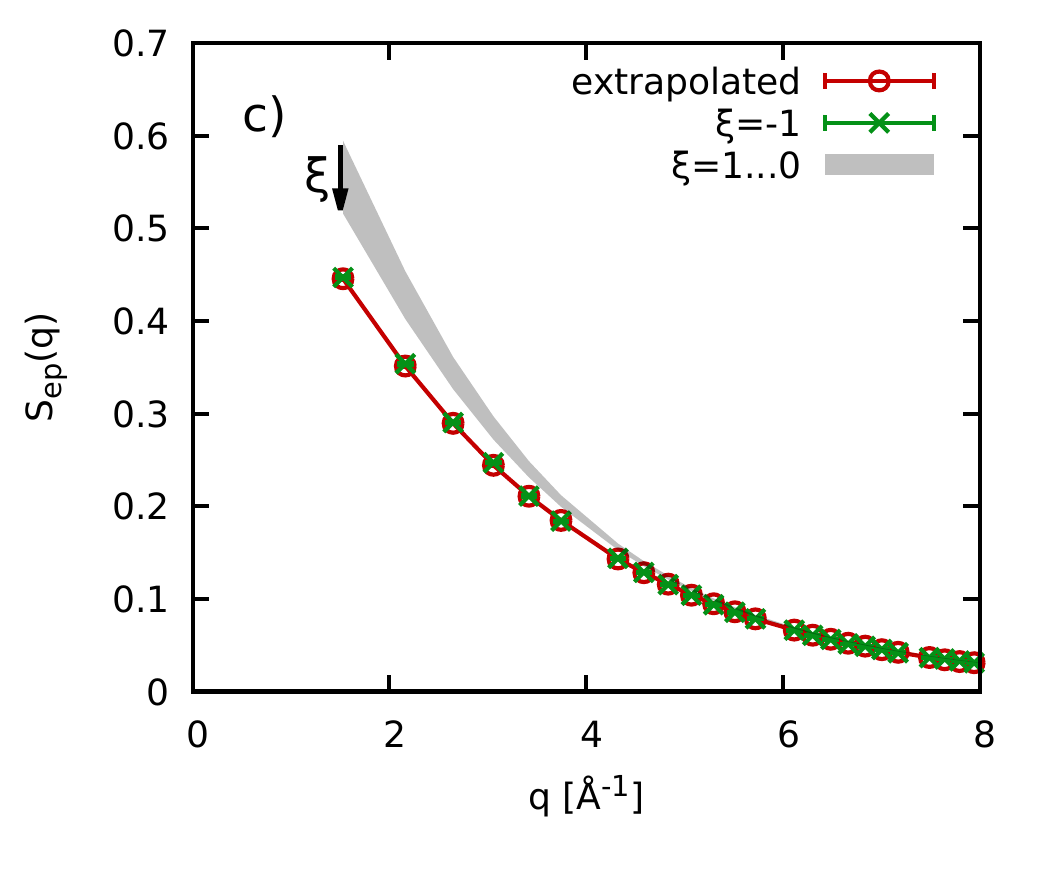}\includegraphics[width=0.45\textwidth]{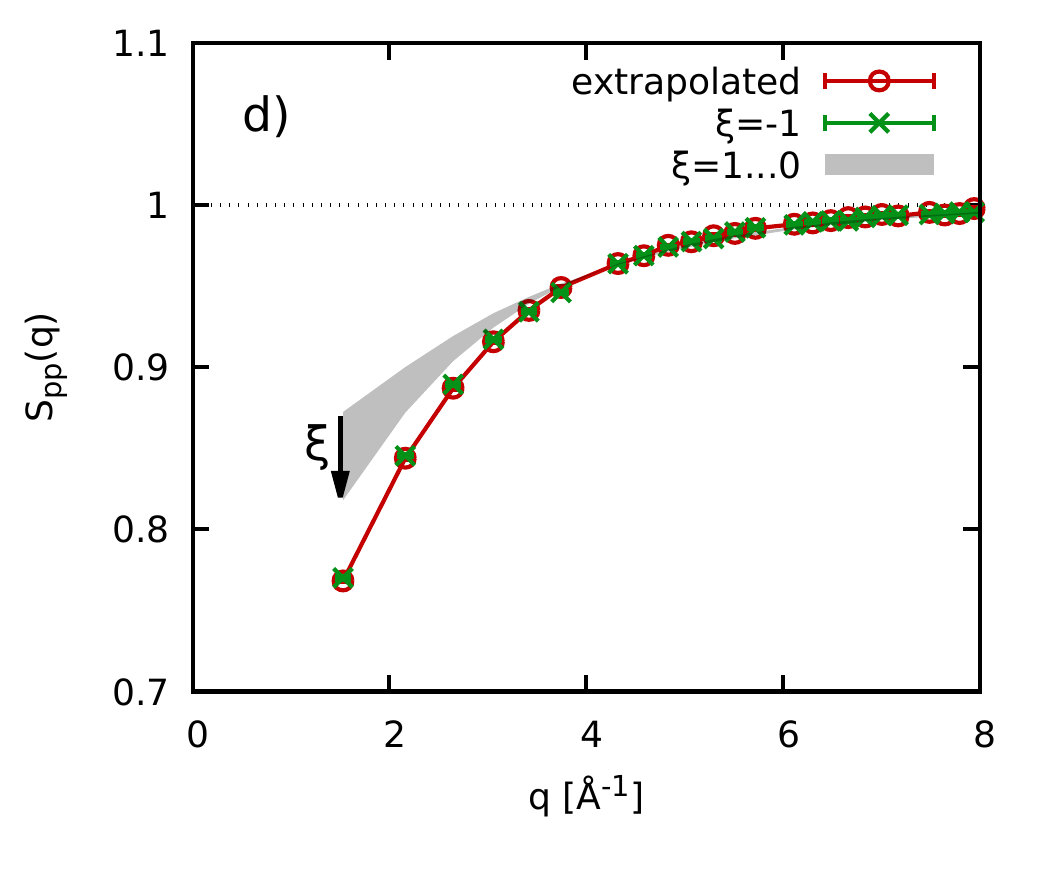}%\\
\caption{\label{fig:Hydrogen_rs2_theta1}  \emph{Ab initio} PIMC results for hydrogen with $N=14$, $r_s=2$, and $\Theta=1$. a) the electron--electron SSF $S_{ee}(\mathbf{q})$, b) thermal electron--electron structure factor $F_{ee}(\mathbf{q},\beta/2)$, c) electron--proton SSF $S_{ep}(\mathbf{q})$, d) proton--proton SSF $S_{pp}(\mathbf{q})$. Green crosses: direct (exact) PIMC results for $\xi=-1$; red circles: $\xi$-extrapolated results [Eq.~(\ref{eq:fit})]; grey area: FSP free domain of $\xi\in[0,1]$.
}
\end{figure*} 

\begin{figure}\centering
\includegraphics[width=0.45\textwidth]{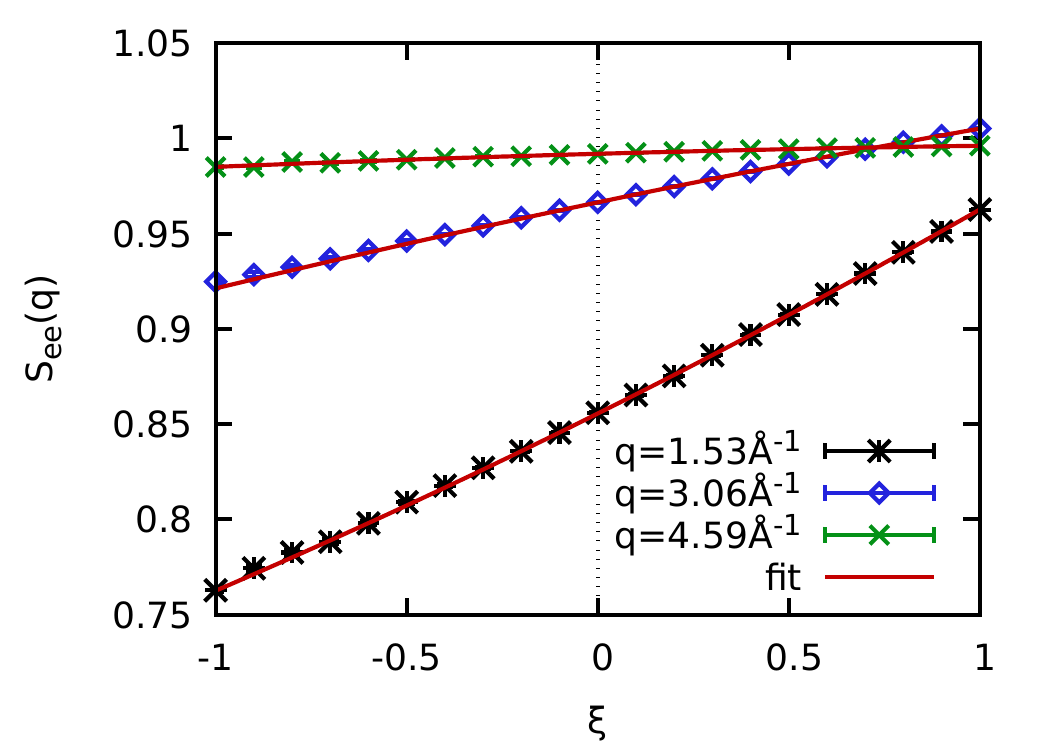}
\caption{\label{fig:Hydrogen_rs2_theta1_xi} The $\xi$-dependence of the electron--electron SSF $S_{ee}(\mathbf{q})$ for three wavenumbers and $N=14$, $r_s=2$, $\Theta=1$. Symbols: PIMC data for $q=1.53\,$\AA$^{-1}$ (black stars), $q=3.06\,$\AA$^{-1}$ (blue diamonds), and $q=4,59\,$\AA$^{-1}$ (green crosses). Solid red lines: fits according to Eq.~(\ref{eq:fit}) based on PIMC data in the FSP free domain of $\xi\in[0,1]$.
}
\end{figure} 

\begin{figure*}\centering
\includegraphics[width=0.45\textwidth]{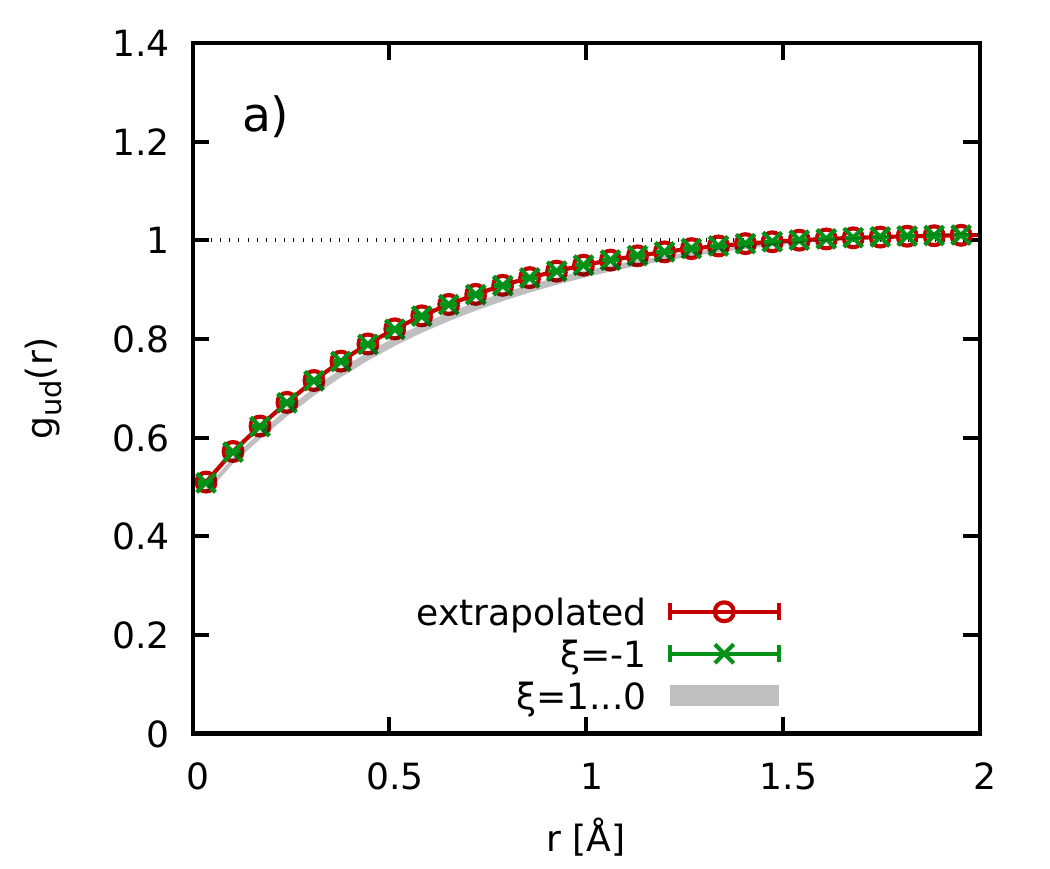}\includegraphics[width=0.45\textwidth]{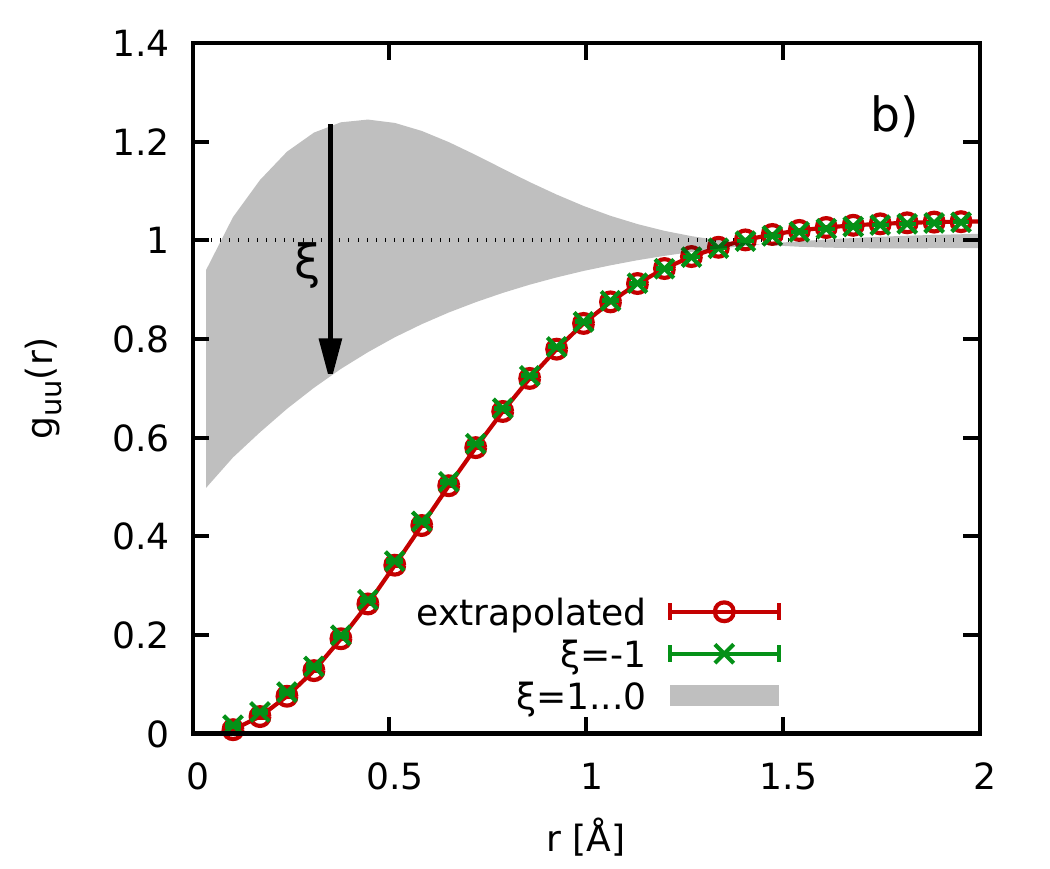}\\\vspace*{-0.92cm}
\includegraphics[width=0.45\textwidth]{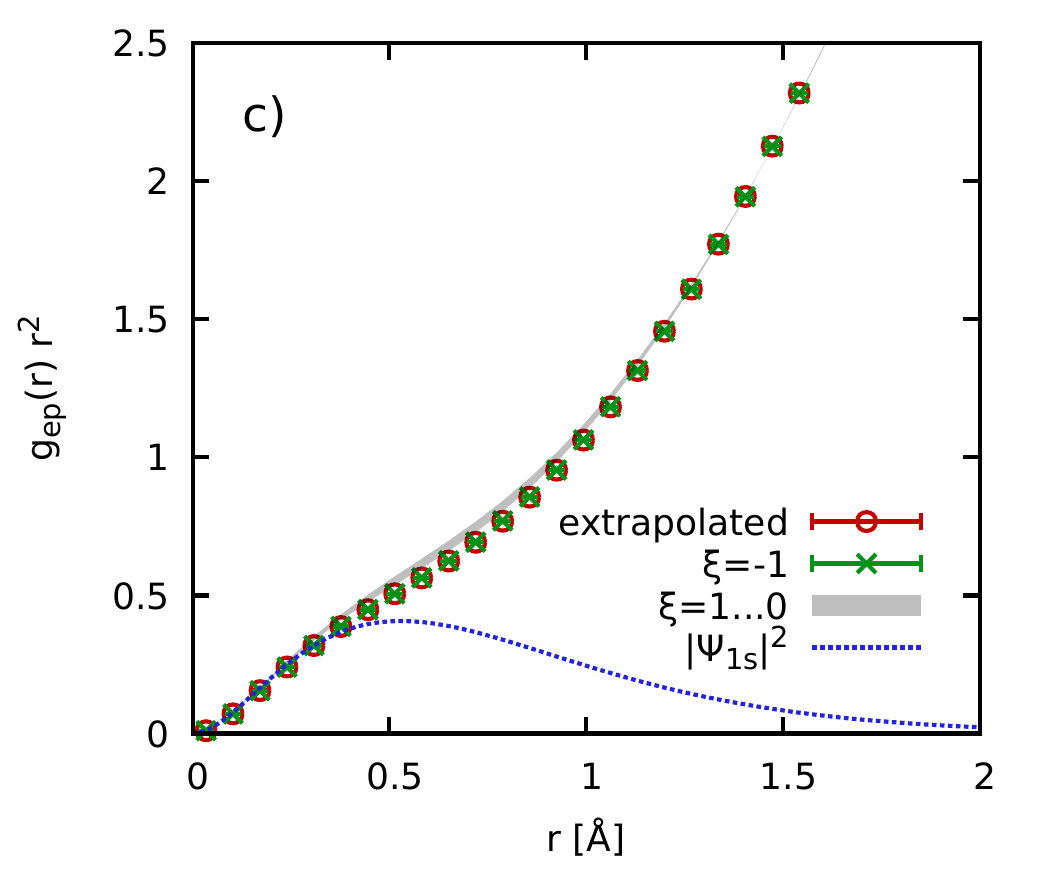}\includegraphics[width=0.45\textwidth]{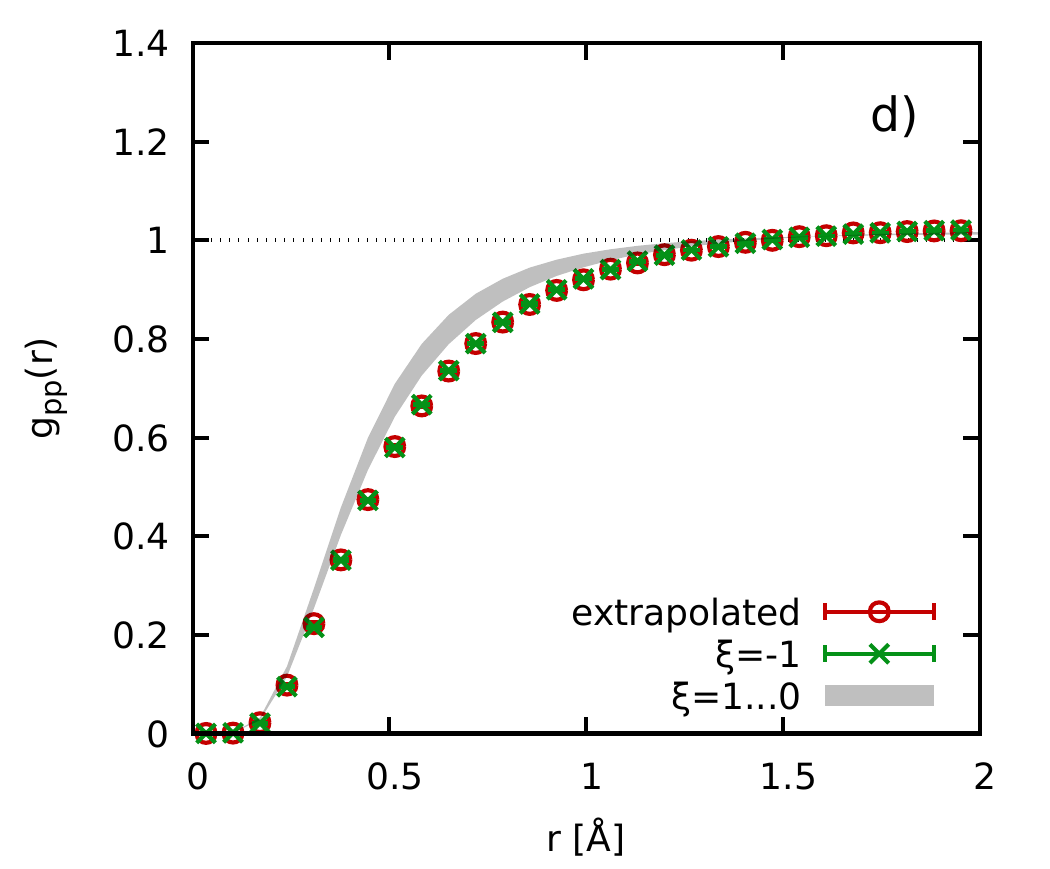}
\caption{\label{fig:Hydrogen_rs2_theta1_PCF}  \emph{Ab initio} PIMC results for hydrogen with $N=14$, $r_s=2$, and $\Theta=1$. a) the spin up--down PCF $g_{ud}(r)$, b) spin up--up PCF $g_{uu}(r)$, c) rescaled electron--proton PCF $g_{ep}(r)$, d) proton--proton PCF $g_{pp}(r)$. Green crosses: direct (exact) PIMC results for $\xi=-1$; red circles: $\xi$-extrapolated results [Eq.~(\ref{eq:fit})]; grey area: FSP free domain of $\xi\in[0,1]$. The dotted blue line is panel c) has been computed from the ground-state wave function of an isolated hydrogen atom, and has been rescaled arbitrarily as a guide to the eye.
}
\end{figure*} 

In Fig.~\ref{fig:Hydrogen_rs2_theta1}, we consider hydrogen at $r_s=2$ ($\rho=0.34\,$g/cc) and $\Theta=1$ ($T=12.53\,$eV). These conditions are at the heart of the WDM regime~\cite{review,wdm_book} and are realized, for example, on the compression path of a fuel capsule in an ICF experiment at the NIF~\cite{Moses_NIF}.
From a physical perspective, hydrogen is expected to be strongly ionized in this regime~\cite{Militzer_PRE_2001,Bohme_PRL_2022,Filinov_PRE_2023}, which means that the electrons can be expected to behave similarly to the UEG at the same conditions.
Given the excellent performance of the $\xi$-extrapolation method for the warm dense UEG~\cite{Dornheim_JCP_2023,Dornheim_JPCL_2024}, this regime constitutes a logical starting point for the present investigation.

In panel a), we show the electron--electron static structure factor (SSF) $S_{ee}(\mathbf{q})$ as a function of the wave number $q$ for $N=14$. The green crosses correspond to direct PIMC results for the fermionic limit of $\xi=-1$; in this case, the simulations are challenging due to the FSP, but still feasible, and we find an average sign~\cite{dornheim_sign_problem} of $S=0.08466(14)$.
The red circles have been computed by fitting Eq.~(\ref{eq:fit}) to the sign-problem free domain of $\xi\in[0,1]$ (shaded grey area) and are in excellent agreement with the exact results for all $q$.
Interestingly, the SSF exhibits a more pronounced structure in the bosonic limit of $\xi=1$, which is likely a consequence of the effective attraction of two identical bosons reported in earlier works~\cite{Dornheim_PRA_2020,Dornheim_NJP_2022}.

In Fig.~\ref{fig:Hydrogen_rs2_theta1_xi}, we show the $\xi$-dependence of $S_{ee}(\mathbf{q})$ for three selected wavenumbers. Specifically, the symbols show our exact direct PIMC results, which are available for all $\xi$ at these parameters, and the solid red lines fits via Eq.~(\ref{eq:fit}) that have been obtained exclusively based on input data from the sign-problem free domain of $\xi\in[0,1]$. The $\xi$-dependence is most pronounced for the smallest depicted $q$-value, and Eq.~(\ref{eq:fit}) nicely reproduces the PIMC results over the entire $\xi$-range in all cases, as it is expected.

In Fig.~\ref{fig:Hydrogen_rs2_theta1}b), we analyze the \emph{thermal structure factor}~\cite{Dornheim_PRR_2022}, defined by the imaginary-time density--density correlation function evaluated at its $\tau=\beta/2$ minimum, i.e., $F_{ee}(\mathbf{q},\beta/2)$. From a theoretical perspective, the ITCF can be expected to depend even more strongly on quantum effects in general, and quantum statistics in particular, and thus constitutes a potentially more challenging case compared to the static $S_{ee}(\mathbf{q})=F_{ee}(\mathbf{q},0)$. Nevertheless, we observe that the $\xi$-extrapolation method is capable of giving excellent results for the thermal structure factor over the entire wavenumber range, just as in the previously investigated case of the UEG~\cite{Dornheim_JCP_2023}.

\begin{figure*}\centering
\includegraphics[width=0.45\textwidth]{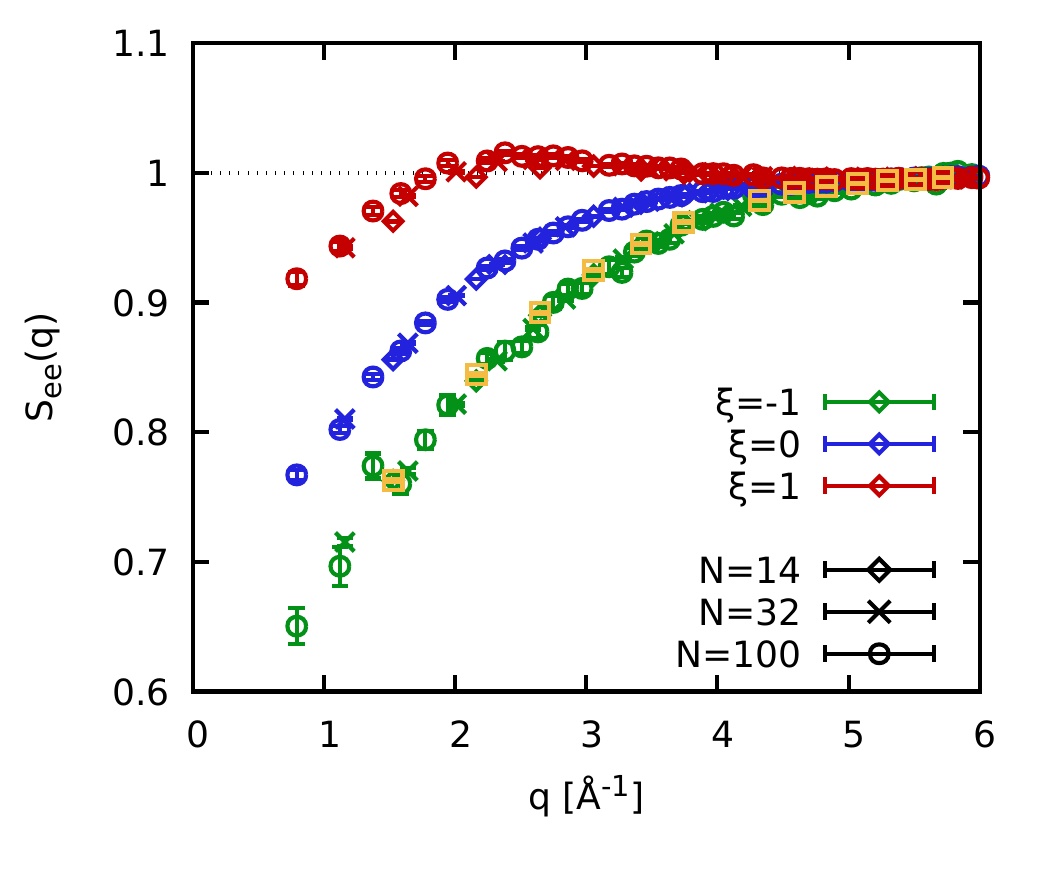}\includegraphics[width=0.45\textwidth]{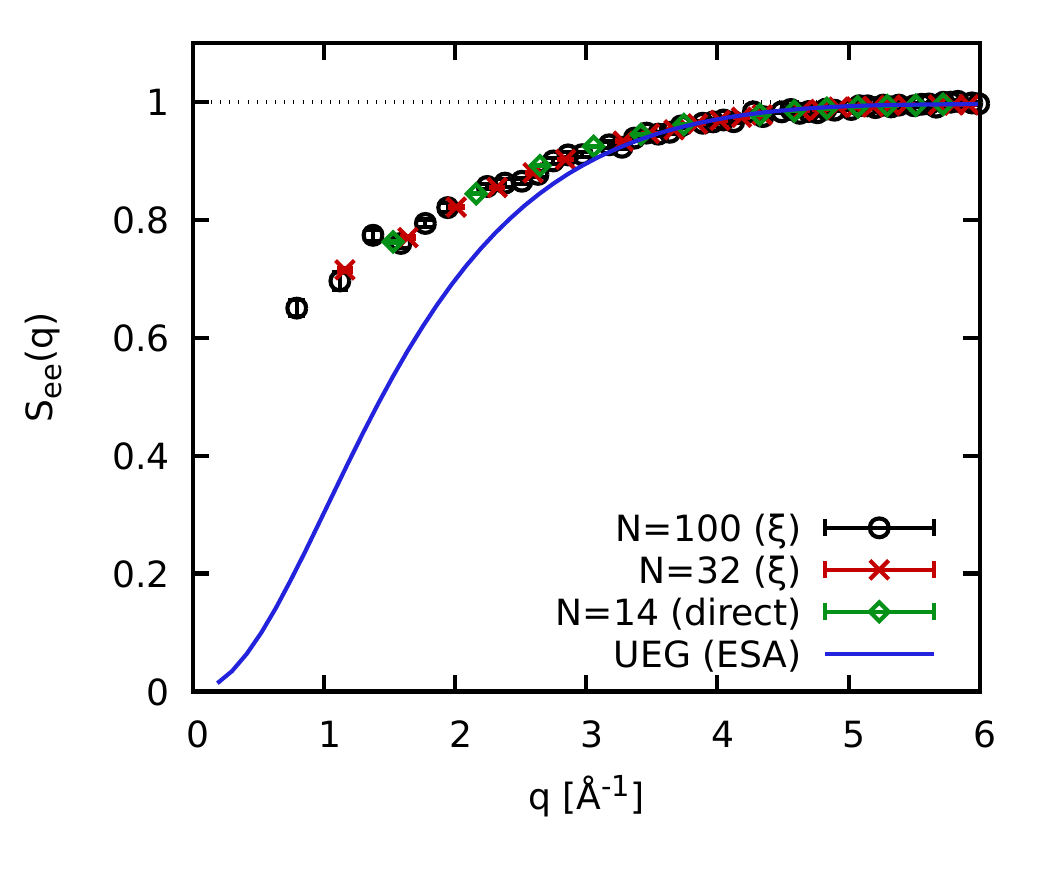}
\caption{\label{fig:Hydrogen_rs2_theta1_N} PIMC results for the electronic static structure factor $S_{ee}(\mathbf{q})$ at $r_s=2$ and $\Theta=1$ for $N=14$ (diamonds), $N=32$ (crosses) and $N=100$ (circles). Left: impact of quantum statistics, with green, blue and red symbols showing results for $\xi=-1$ (extrapolated), $\xi=0$, and $\xi=1$; the yellow squares show direct PIMC results for $N=14$ and $\xi=-1$. Right: comparing PIMC results for hydrogen to UEG results at the same parameters computed via the ESA~\cite{Dornheim_PRL_2020_ESA,Dornheim_PRB_ESA_2021}.
}
\end{figure*} 

Let us next focus more explicitly on the effect of the nuclei (i.e., protons). To this end, we show the electron--proton SSF $S_{ep}(\mathbf{q})$ and proton--proton SSF $S_{pp}(\mathbf{q})$ in panels c) and d). First and foremost, we find the same excellent agreement between the exact, direct PIMC results and the $\xi$-extrapolation results in both cases for all $q$. Second, the impact of quantum statistics is comparably reduced in particular for $S_{pp}(\mathbf{q})$, which is very similar to $S_{ee}(\mathbf{q})$ in the fermionic limit, but qualitatively differs substantially in the bosonic limit of $\xi=1$. 
Such direct insights into the importance of quantum degeneracy effects on different observables are a nice side effect of the $\xi$-extrapolation method.

To focus more closely on such effects, we investigate the spin-resolved electron--electron pair correlation functions (PCFs) $g_{ud}(r)$ and $g_{uu}(r)$ in Fig.~\ref{fig:Hydrogen_rs2_theta1_PCF}a) and b), respectively. In the spin-offdiagonal case, hardly any effects of quantum statistics can be resolved. While somewhat expected, this is still different from $S_{pp}(\mathbf{q})$, for which quantum statistics prove to be important even though the protons themselves are effectively distinguishable, see Sec.~\ref{sec:PIMC} above.
In stark contrast, the behaviour of the spin-diagonal PCF is predominantly shaped by quantum statistics, in particular for small separations $r\to0$. As mentioned above, bosons tend to cluster around each other, resulting in a large contact probability $g_{uu}^{\xi=1}(0)$, and a nontrivial maximum around $r=0.4\,$\AA. For distinguishable Boltzmann quantum particles, there is still a finite contact probability of $g_{uu}^{\xi=0}(0)\approx0.5$, but the bosonic maximum for small but finite $r$ disappears. In the fermionic limit, the contact probability completely vanishes due to the Pauli exclusion principle. It is striking that the $\xi$-extrapolation method very accurately captures these stark qualitative differences for all $r$, and the red circles nicely recover the exact direct PIMC results that are shown by the green crosses. For completeness, we note that the on-top PCF $g(0)$ constitutes a very important property even for the UEG~\cite{Hunger_PRE_2021,Dornheim_PRL_2020_ESA,Dornheim_PRB_ESA_2021,tolias2024_VS}. The applicability of the $\xi$-extrapolation method thus opens up the possibility to study $g_{ee}(0)$ and to reduce finite-size effects~\cite{Hunger_PRE_2021} and the possible impact of nodal errors in the case of restricted PIMC~\cite{Brown_PRL_2013} by simulating larger system sizes as demonstrated in the recent Ref.~\cite{Dornheim_JPCL_2024}.

In Fig.~\ref{fig:Hydrogen_rs2_theta1_PCF}c) we show the electron--proton PCF $g_{ep}(r)$ [rescaled by a factor of $r^2$], which contains information about the electronic localization around the protons. In an atomic system, one would expect a maximum around the Bohr radius of $r=0.529\,$\AA~\cite{Filinov_PRE_2023}, see the dotted blue curve showing the probability density of an electron around an isolated proton at the hydrogen ground state.
For this observable, quantum statistics effects are mostly restricted around intermediate distances, while the contact range between an electron and a proton is mostly unaffected. 
Finally, we show the proton--proton PCF $g_{pp}(r)$ in Fig.~\ref{fig:Hydrogen_rs2_theta1_PCF}d). It is largely featureless, except for a pronounced exchange--correlation hole around $r=0$. Interestingly, fermionic exchange effects play a more prominent role compared to $g_{ep}(r)$, and they are perfectly captured by the $\xi$-extrapolation at these conditions.

\begin{figure*}\centering
\includegraphics[width=0.45\textwidth]{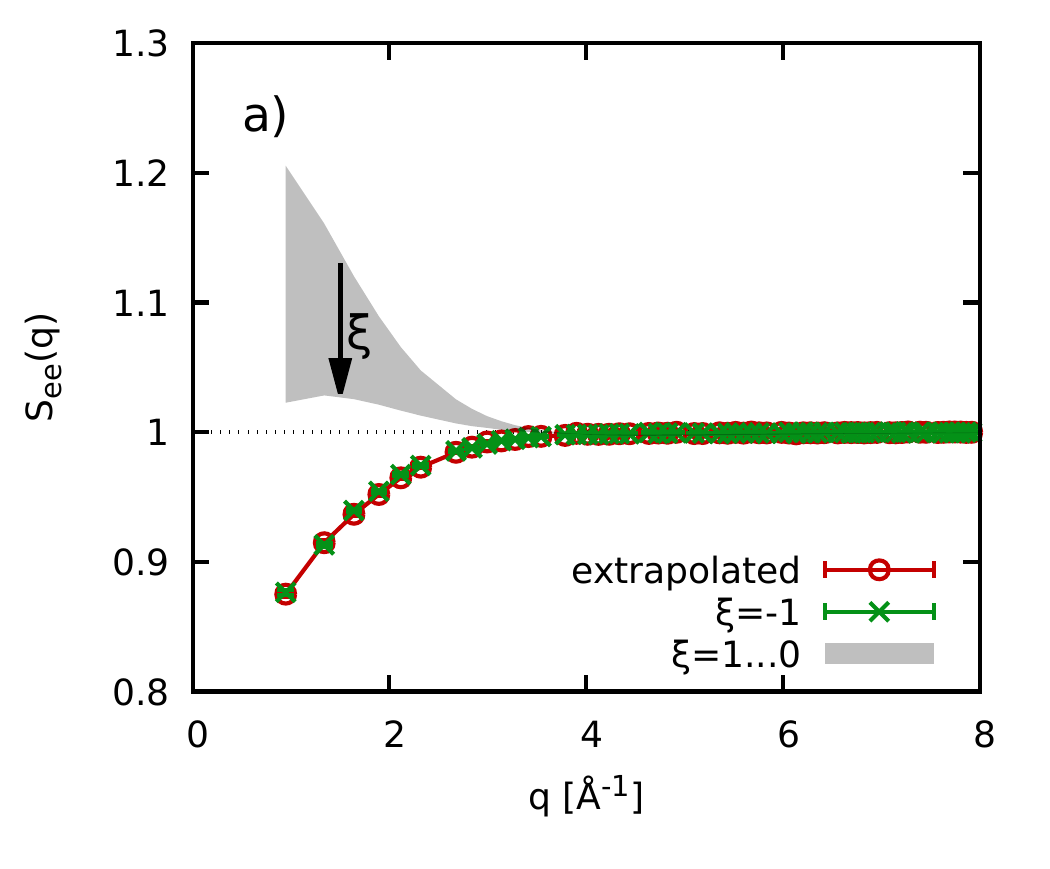}\includegraphics[width=0.45\textwidth]{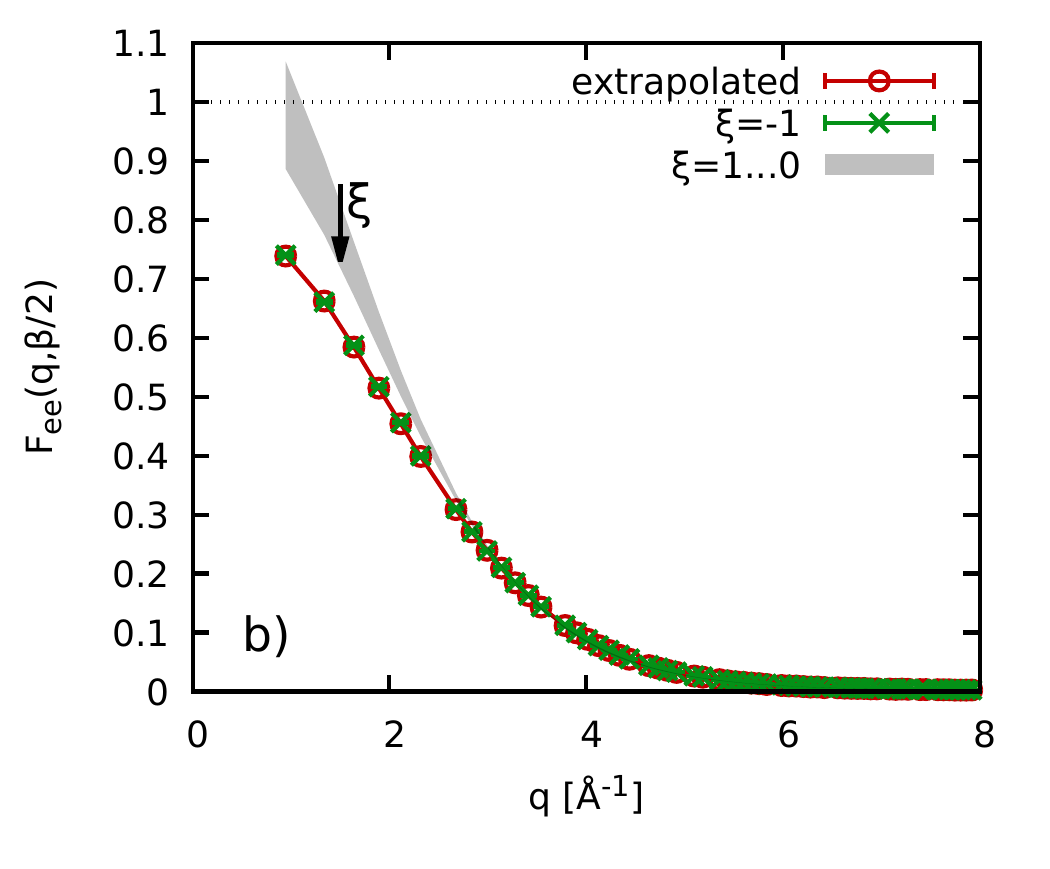}\\\vspace*{-1.25cm}
\includegraphics[width=0.45\textwidth]{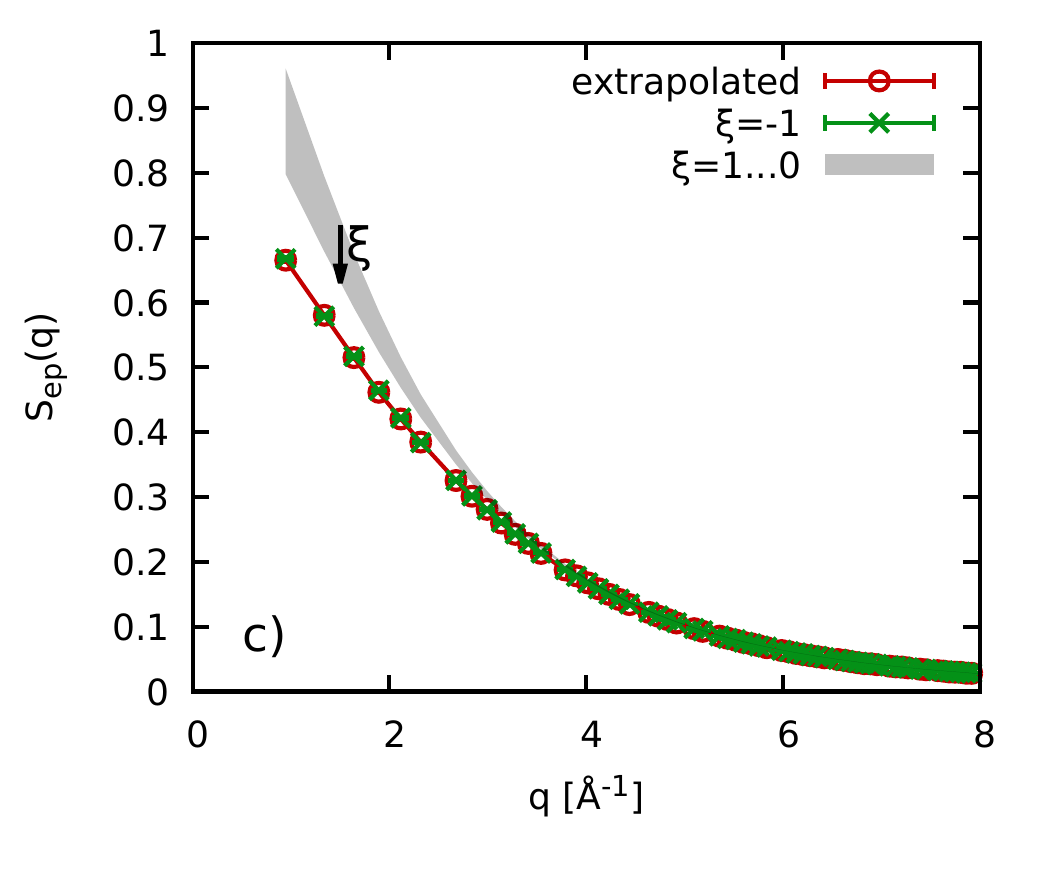}\includegraphics[width=0.45\textwidth]{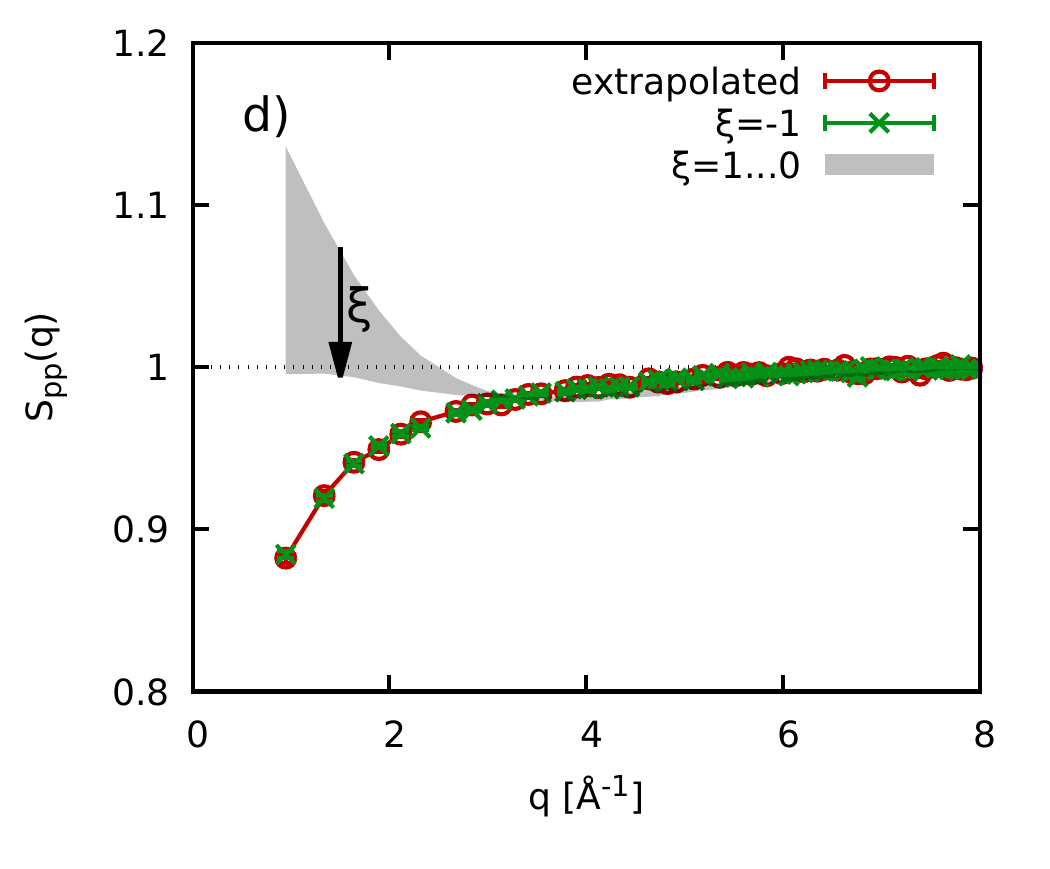}
\caption{\label{fig:Hydrogen_rs3p23_theta1}  \emph{Ab initio} PIMC results for hydrogen with $N=14$, $r_s=3.23$, and $\Theta=1$. a) electron--electron SSF $S_{ee}(\mathbf{q})$, b) thermal electron--electron structure factor $F_{ee}(\mathbf{q},\beta/2)$, c) electron--proton SSF $S_{ep}(\mathbf{q})$, d) proton--proton SSF $S_{pp}(\mathbf{q})$. Green crosses: direct (exact) PIMC results for $\xi=-1$; red circles: $\xi$-extrapolated results [Eq.~(\ref{eq:fit})]; grey area: FSP free domain of $\xi\in[0,1]$. %\textcolor{red}{TD: better statistics for $S_{pp}(q)$ still evaluating...}
}
\end{figure*} 

Let us conclude our analysis of the metallic density case with an investigation of finite-size effects. To this end, we show the electronic SSF $S_{ee}(\mathbf{q})$ in Fig.~\ref{fig:Hydrogen_rs2_theta1_N} for $N=14$ (diamonds), $N=32$ (crosses), and $N=100$ (circles) hydrogen atoms. The red, blue, and green symbols in the left panel correspond to $\xi=1$, $\xi=0$, and $\xi=-1$ (extrapolated), respectively; the yellow squares show the corresponding direct PIMC results for $\xi=-1$ which are available only for $N=14$. 
Apparently, no dependence on the system size can be resolved for the fermionic limit of prime interest for the present work. This is consistent both with previous findings for the UEG model~\cite{dornheim_prl,review,Dornheim_JCP_2021,Holzmann_PRB_2016,Chiesa_PRL_2006,Drummond_PRB_2008}, and also with the well-known principle of electronic nearsightedness~\cite{Kohn_PNAS_2005}.
Similarly, we cannot resolve any clear dependence on $N$ in the case of $\xi=0$ despite the smaller error bars.
Interestingly, this situation somewhat changes for the bosonic case of $\xi=1$, where e.g.~the results for $N=14$ exhibit small differences compared to the other data.
Heuristically, this can be attributed to a breakdown of nearsightedness for bosons, which, in the extreme case, are known for exhibiting off-diagonal long-range order e.g.~in the case of superfluidity~\cite{PhysRevB.72.014533}.
It is important to note that the $\xi$-extrapolation still gives the correct fermionic limit despite the larger finite-size effects in the $\xi>0$ data, as it becomes evident from the excellent agreement between the green diamonds and yellow squares.

The right panel of Fig.~\ref{fig:Hydrogen_rs2_theta1_N} shows a comparison of our new PIMC results for hydrogen with the UEG model at the same conditions; the latter has been computed from the \emph{effective static approximation} (ESA)~\cite{Dornheim_PRL_2020_ESA,Dornheim_PRB_ESA_2021}, which is known to be quasi-exact in this regime.
For large $q$, $S_{ee}(\mathbf{q})$ approaches the single-particle limit for both hydrogen and the UEG, and they agree for $q\gtrsim3\,$\AA$^{-1}$. 
In the long wavelength limit of $q\to0$, $S_{ee}(\mathbf{q})$ is described by a parabola vanishing for $q=0$; this is a well-known consequence of perfect screening in the UEG~\cite{review,kugler_bounds,quantum_theory}.
In contrast, $S_{ee}(\mathbf{q})$ attains a finite value for real electron--ion systems that is governed by the compressibility sum-rule~\cite{CSR}.
This can be explained by considering the definition of $S_{ee}(\mathbf{q})$ as the normalization of the dynamic structure factor,
\begin{eqnarray}\label{eq:DSF}
    S_{ee}(\mathbf{q}) = \int_{-\infty}^\infty \textnormal{d}\omega\ S_{ee}(\mathbf{q},\omega)\ .
\end{eqnarray}
For the UEG, $S_{ee}(\mathbf{q},\omega)$ consists of a single (collective) plasmon peak for small $q$. For hydrogen, this \emph{free electron gas} feature is complemented by a) a contribution due to transitions between bound and free states~\cite{boehme2023evidence} and b), more importantly in this context, a quasi-elastic feature due to effectively \emph{bound} electrons and the screening cloud of free electrons~\cite{Vorberger_PRE_2015,siegfried_review}.
While the plasmon weight vanishes for small $q$, this trend does not hold for the other contributions in the case of hydrogen, leading to a finite value of $S_{ee}(0)$.

\subsection{Hydrogen: solid density\label{sec:H_solid}}

\begin{figure*}\centering
\includegraphics[width=0.45\textwidth]{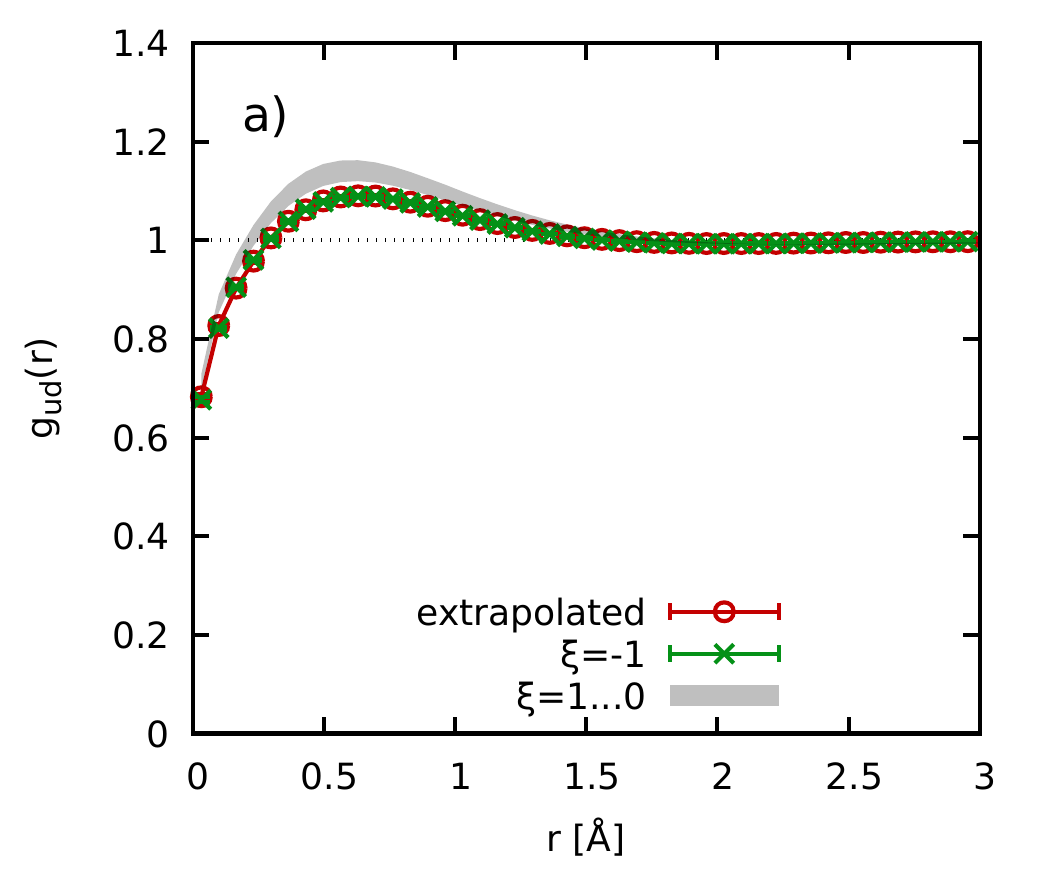}\includegraphics[width=0.45\textwidth]{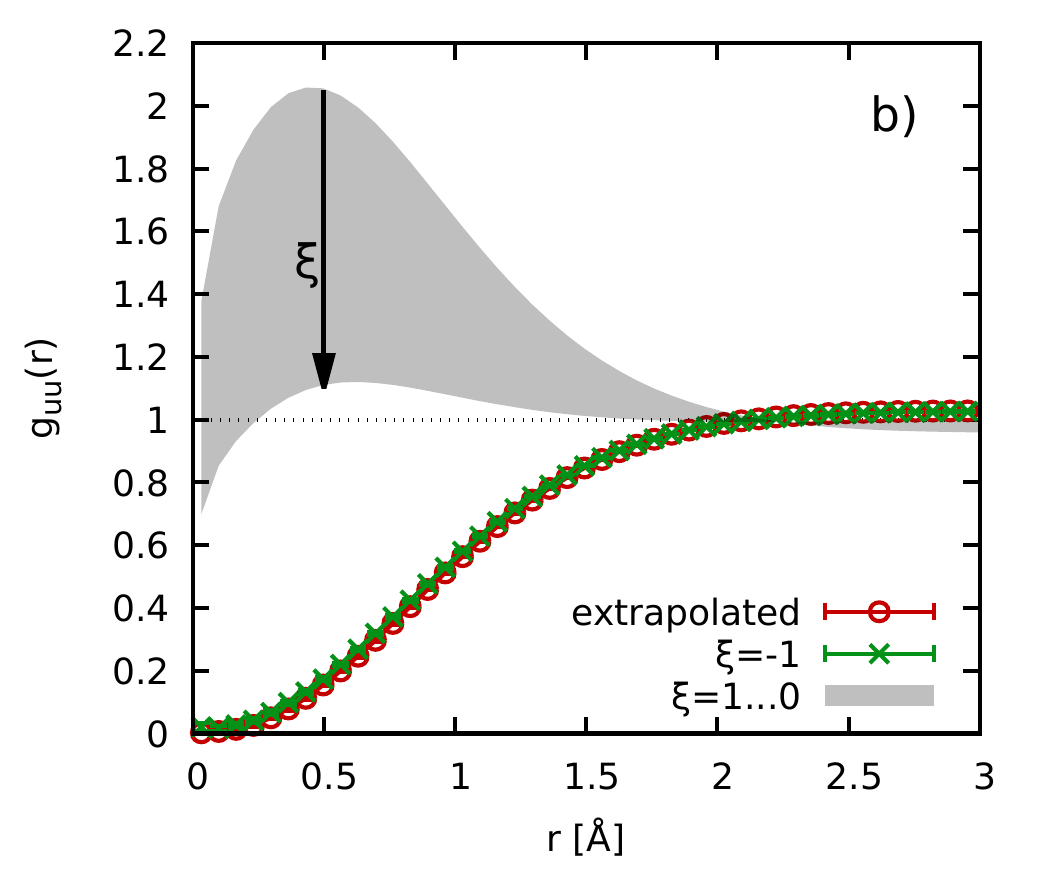}\\\vspace*{-0.92cm}
\includegraphics[width=0.45\textwidth]{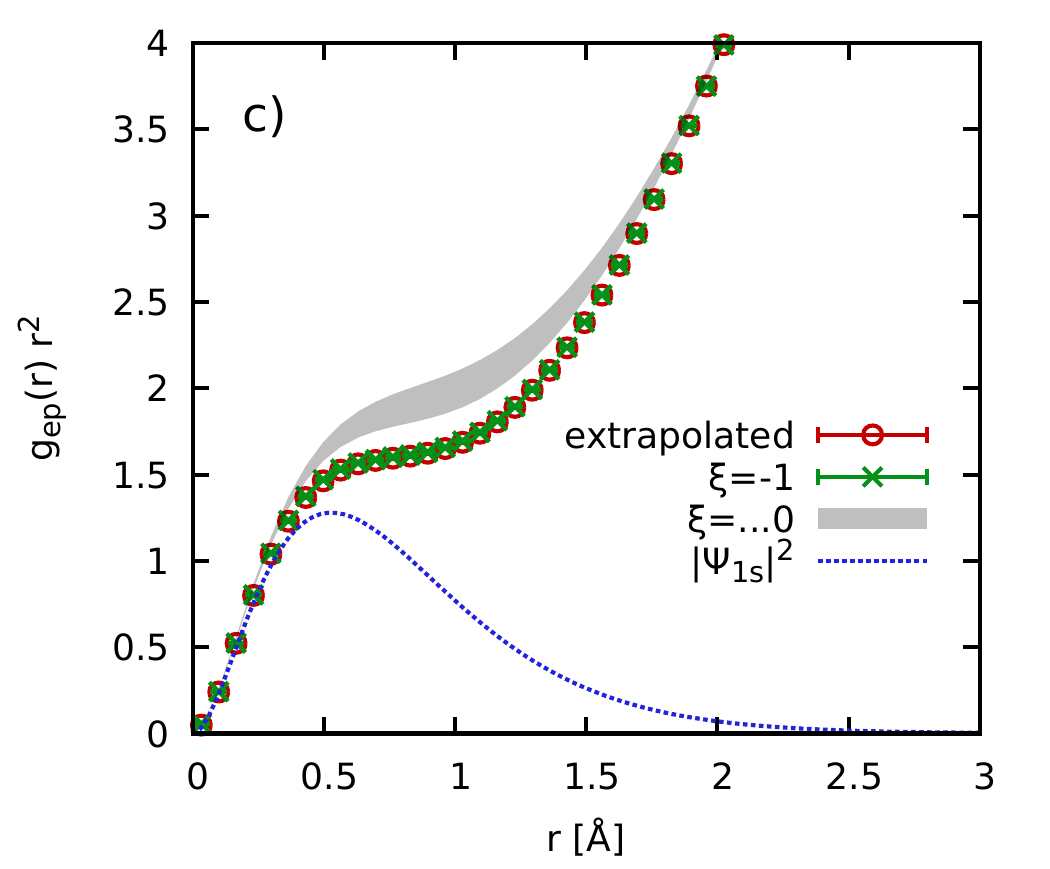}\includegraphics[width=0.45\textwidth]{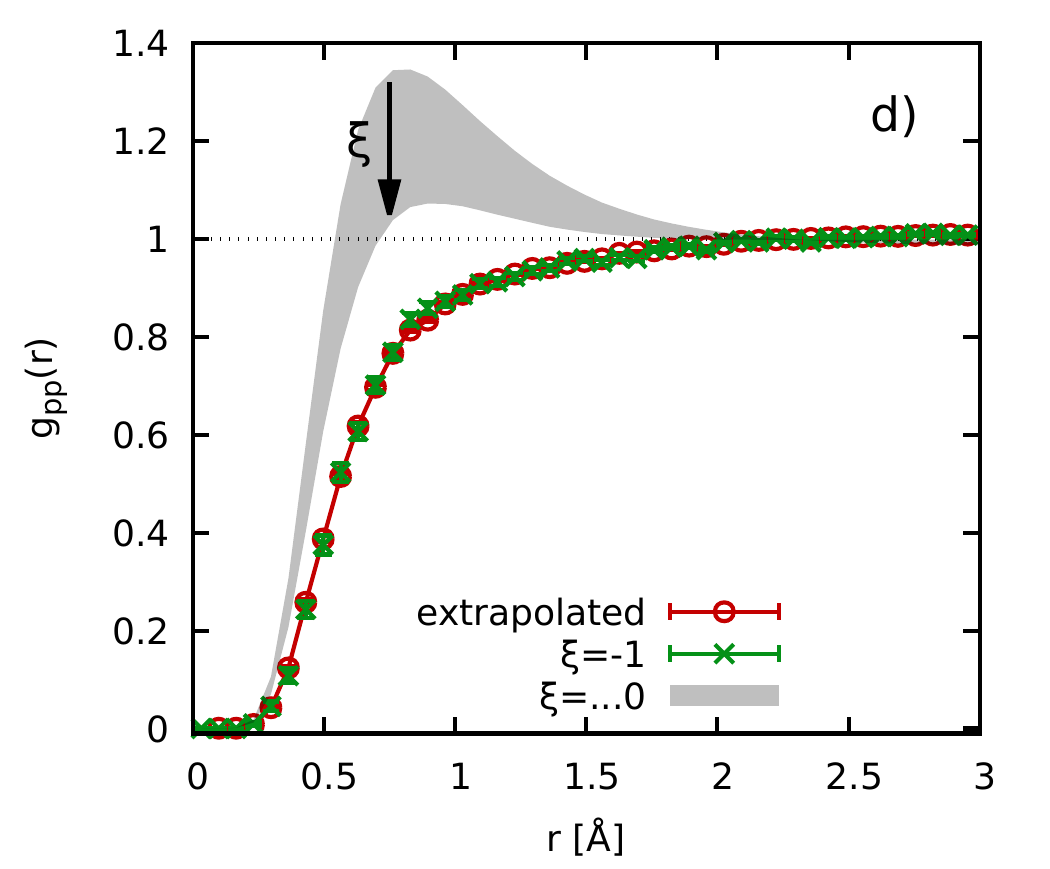}
\caption{\label{fig:Hydrogen_rs3p23_theta1_PCF}  \emph{Ab initio} PIMC results for hydrogen with $N=14$, $r_s=3.23$, and $\Theta=1$. a) spin up--down PCF $g_{ud}(r)$, b) spin up--up PCF $g_{uu}(r)$, c) electron--proton PCF $g_{ep}(r)$, d) proton--proton PCF $g_{pp}(r)$. Green crosses: direct (exact) PIMC results for $\xi=-1$; red circles: $\xi$-extrapolated results [Eq.~(\ref{eq:fit})]; grey area: FSP free domain of $\xi\in[0,1]$. The dotted blue line is panel c) has been computed from the ground-state wave function of an isolated hydrogen atom, and has been rescaled arbitrarily as a guide to the eye.
}
\end{figure*}

Let us next consider hydrogen at solid density, i.e., $r_s=3.23$ ($\rho=0.08\,$g/cc) at $\Theta=1$ ($T=4.80\,$eV). Such conditions can be realized e.g.~in experiments with hydrogen jets~\cite{Zastrau} which can be optically heated and subsequently be probed with XRTS~\cite{Fletcher_Frontiers_2022}. From a physical perspective, such conditions may give rise to interesting effects such as a non-monotonic dispersion relation of the dynamic structure factor at intermediate wave numbers~\cite{Hamann_PRR_2023,Dornheim_Nature_2022} resembling the \emph{roton feature} known e.g.~from ultracold helium~\cite{Godfrin2012,Trigger,Dornheim_SciRep_2022,Ferre_PRB_2016}. From a technical perspective, lower densities are, generally, more challenging for theoretical methods due to the increased impact of XC-effects and the larger degree of inhomogeneity~\cite{low_density1,low_density2}. In the case of PIMC, the same trend holds w.r.t~the FSP as it has been explained during the discussion of Fig.~\ref{fig:Hydrogen_Sign_theta1} above.

In Fig.~\ref{fig:Hydrogen_rs3p23_theta1}, we show extensive new PIMC results for the $\xi$-extrapolation of various structural properties at $\Theta=1$. Panel a) corresponds to $S_{ee}(q)$ and is flat to a high degree. This is mainly a consequence of the electronic localization around protons. At the same time, we find that the $\xi$-extrapolation works with very high accuracy and reproduces the exact direct PIMC results for $\xi=-1$ for all $q$ within the given level of accuracy. The same holds for the thermal structure factor $F_{ee}(\mathbf{q},\beta/2)$, electron--proton SSF $S_{ep}(\mathbf{q})$, and proton--proton SSF $S_{pp}(\mathbf{q})$ shown in panels b), c), and d), respectively.

\begin{figure*}\centering
\includegraphics[width=0.45\textwidth]{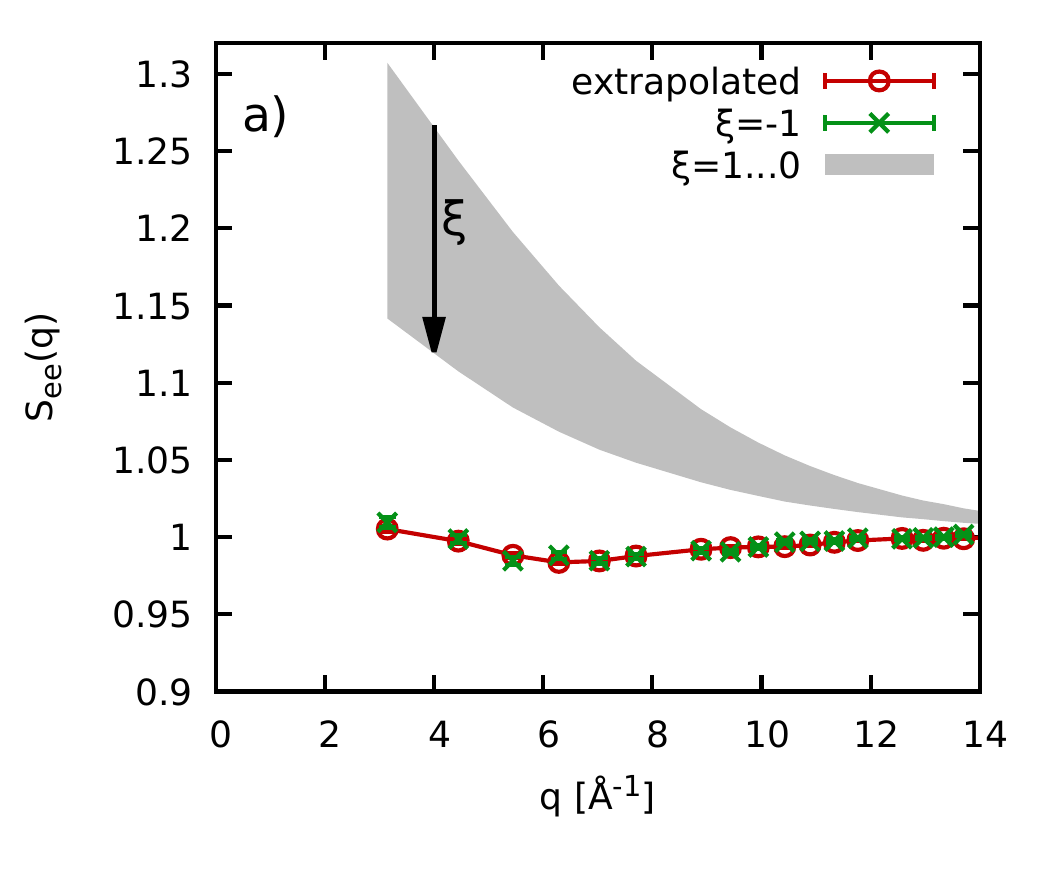}\includegraphics[width=0.45\textwidth]{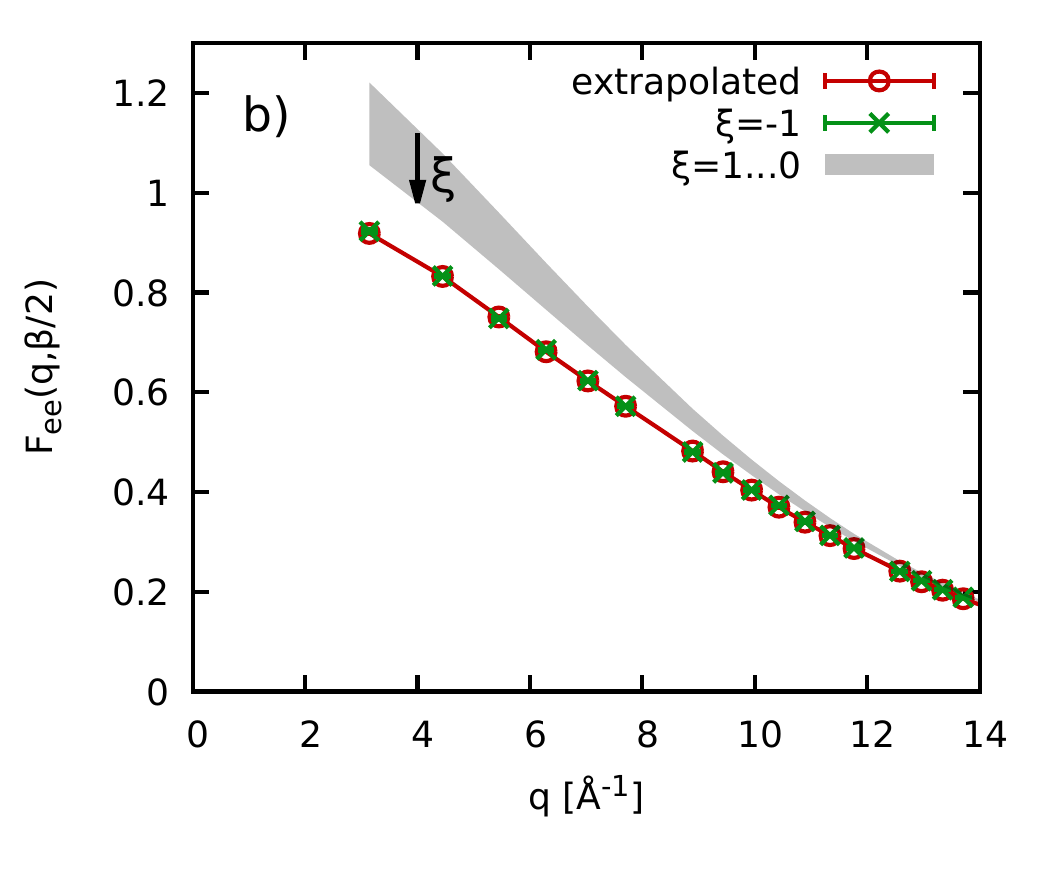}\\\includegraphics[width=0.45\textwidth]{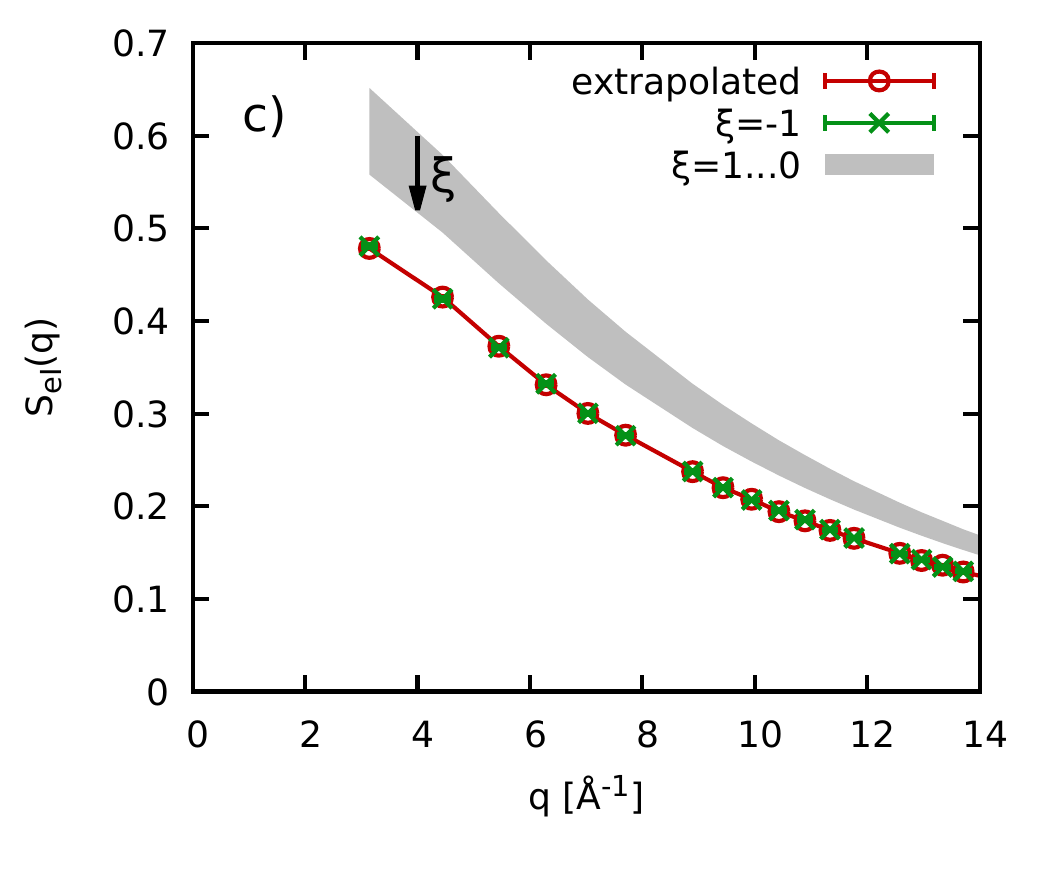}\includegraphics[width=0.45\textwidth]{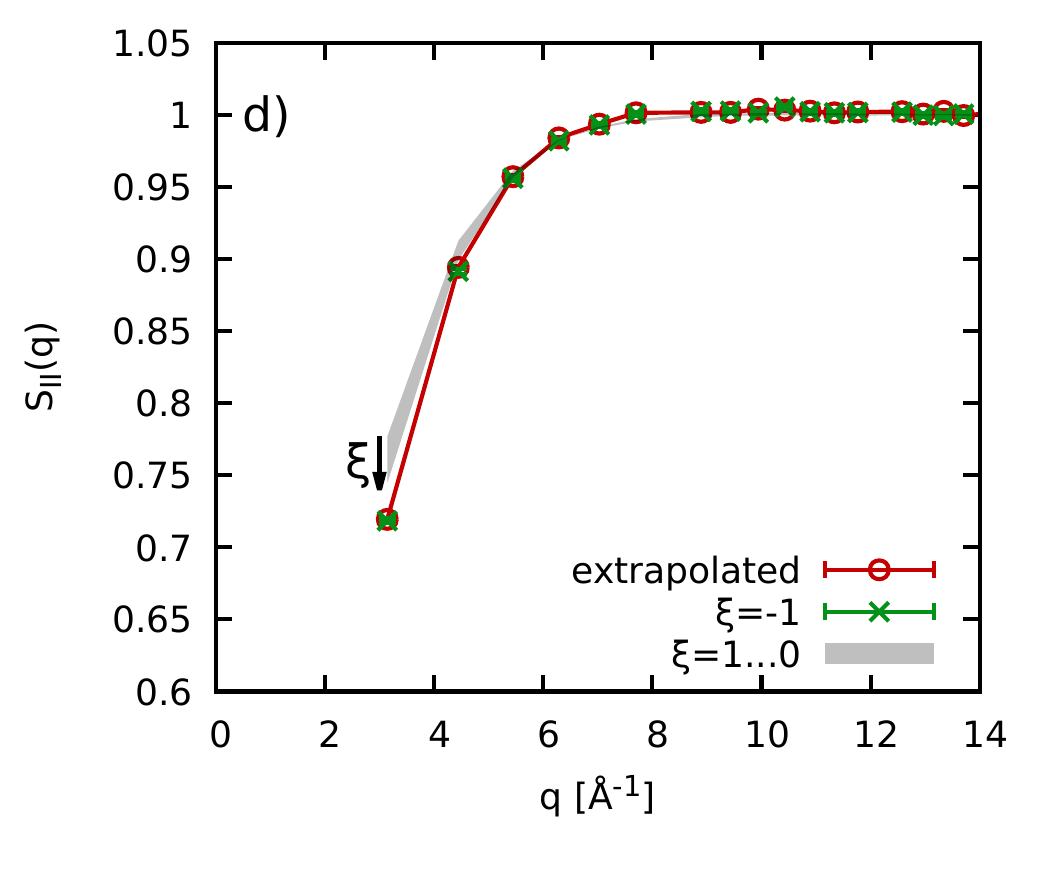}\\
\includegraphics[width=0.45\textwidth]{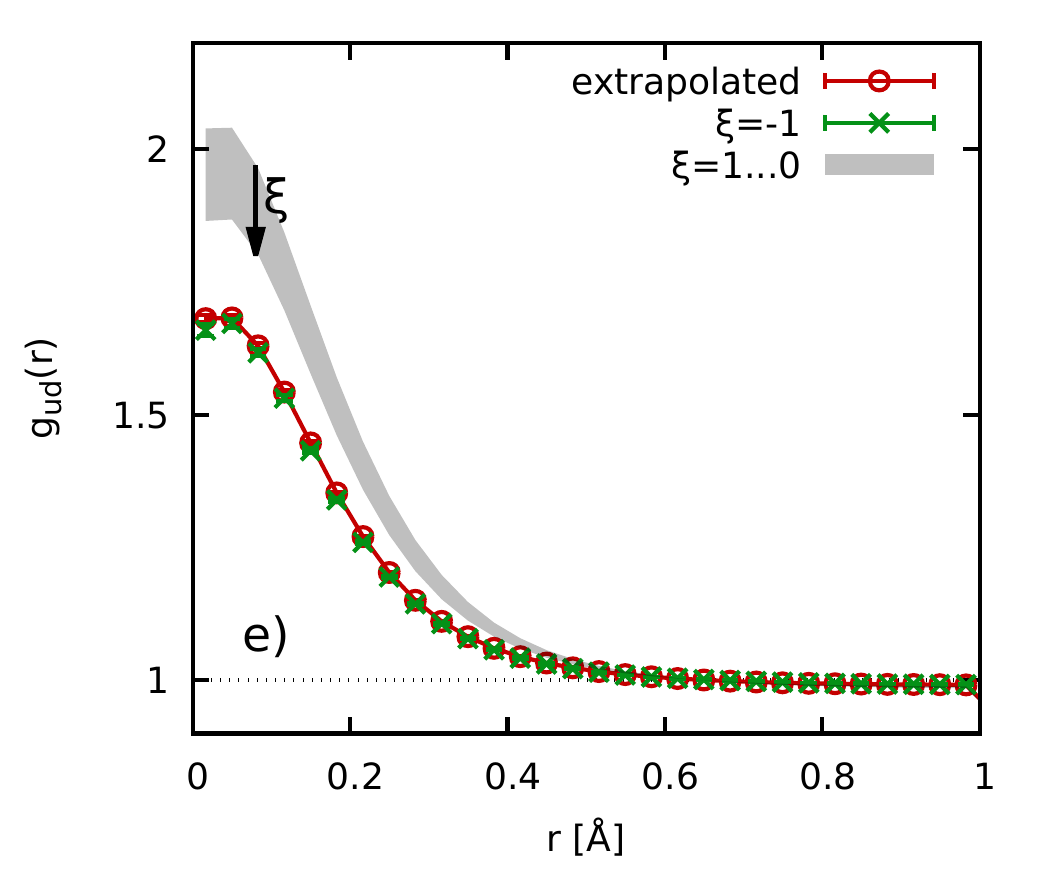}\includegraphics[width=0.45\textwidth]{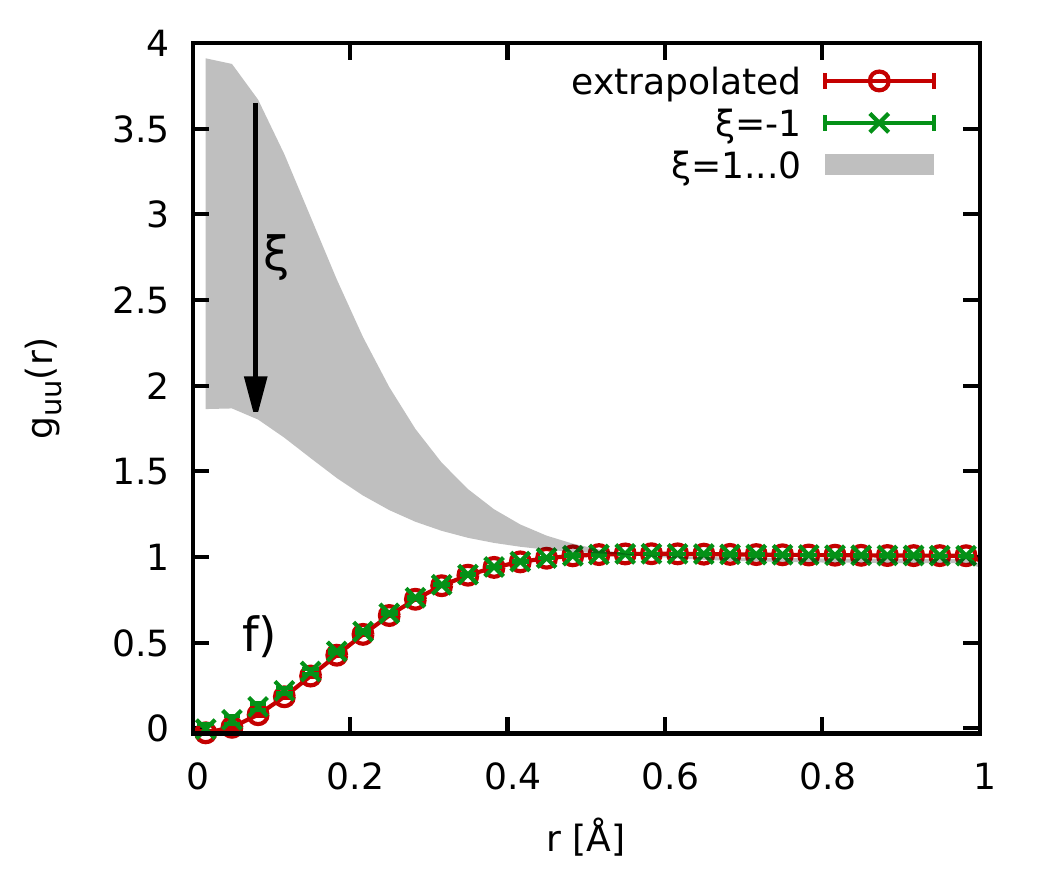}
\caption{\label{fig:Be_rs0p93_T100eV_N4}  \emph{Ab initio} PIMC results for Be with $r_s=0.93$ ($\rho=7.5\,$g/cc), $\Theta=1.73$ ($T=100\,$eV), and $N=4$. a) electron--electron SSF $S_{ee}(\mathbf{q})$, b) thermal electron--electron structure factor $F_{ee}(\mathbf{q},\beta/2)$, c) electron--proton SSF $S_{ep}(\mathbf{q})$, d) proton--proton SSF $S_{pp}(\mathbf{q})$, e) spin up--down PCF $g_{ud}(r)$, f) spin up--up PCF $g_{uu}(r)$. Green crosses: direct (exact) PIMC results for $\xi=-1$; red circles: $\xi$-extrapolated results [Eq.~(\ref{eq:fit})]; grey area: FSP free domain of $\xi\in[0,1]$. 
}
\end{figure*} 

In Fig.~\ref{fig:Hydrogen_rs3p23_theta1_PCF}a), we show the spin-offdiagonal PCF $g_{ud}(r)$. First, we find a somewhat more pronounced impact of quantum statistics in the solid density case compared to $r_s=2$ that has been investigated in Fig.~\ref{fig:Hydrogen_rs2_theta1} above. Second, $g_{ud}(r)$ exhibits an interesting and nontrivial structure with a significant maximum around $r=0.5\,$\AA. It is a sign of the formation of H$^{-}$ ions and the incipient formation of molecules in the system. Third, the $\xi$-extrapolation method again works well for all $r$. The corresponding spin-diagonal PCF $g_{uu}(r)$ is shown in Fig.~\ref{fig:Hydrogen_rs3p23_theta1_PCF}b) and exhibits a strikingly different behaviour. In this case, the impact of quantum statistics is approximately $100\%$, and we find a pronounced \emph{exchange--correlation hill} around $r=0.5\,$\AA~for $\xi=1$. Nevertheless, the $\xi$-extrapolation method accurately captures the correct XC-hole in the fermionic limit. We note that it holds $g_{ud}(r)=g_{uu}(r)$ for $\xi=0$.
From a physical perspective, the bosonic effective attraction leads to a clustering of spin-aligned electrons and, in this way, to the formation of bosonic molecules. This can be seen particularly well in Fig.~\ref{fig:Hydrogen_rs3p23_theta1_PCF}d), where we show the proton--proton PCF $g_{pp}(r)$. It exhibits a pronounced peak around the molecular distance of $r=0.74\,$\AA~\cite{Moldabekov_JCTC_BoundState_2023} that vanishes with decreasing $\xi$. It is entirely absent for $\xi=-1$, which is fully captured by the $\xi$-extrapolation method. 
Finally, panel Fig.~\ref{fig:Hydrogen_rs3p23_theta1_PCF}c) shows the electron--proton PCF $g_{ep}(r)$. While it is qualitatively similar to the case of $r_s=2$ shown in Fig.~\ref{fig:Hydrogen_rs2_theta1_PCF} above, it exhibits a nearly flat progression for $a_\textnormal{B}\lesssim r \lesssim 2a_\textnormal{B}$; this feature is indicative of a somewhat lower degree of electronic delocalization, as it is expected.

%Finally, Fig.~\ref{fig:Hydrogen_rs3p23_theta1}f), we show the spin-diagonal electronic PCF $g_{uu}(r)$

%\begin{figure}\centering
%\includegraphics[width=0.45\textwidth]{Hydrogen_N14_rs3.23_theta1_proton_xi_extrapolation.pdf}
%\caption{\label{fig:Hydrogen_rs3p23_theta1_proton} \emph{Ab inito} PIMC results for the proton--proton pair correlation function $g_{pp}(r)$ for $N=14$, $r_s=3.23$, and $\Theta=1$. \textcolor{red}{To be removed!}
%}
%\end{figure} 

%In Fig.~\ref{fig:Hydrogen_rs3p23_theta1_proton}, we analyze potential finite-size effects in $S_{ee}(\mathbf{q})$. 
%The 

%\begin{figure}\centering
%\includegraphics[width=0.45\textwidth]{FSC_Hydrogen_SSF_rs3.23_theta1_N.pdf}\\\vspace*{-1.25cm}
%\includegraphics[width=0.45\textwidth]{Hydrogen_SSF_rs3.23_theta1_N.pdf}
%\caption{\label{fig:Hydrogen_rs3p23_theta1_N} PIMC results for the electronic static structure factor $S_{ee}(\mathbf{q})$ at $r_s=3.23$ and $\Theta=1$ for $N=14$ (diamonds), $N=32$ (crosses) and $N=100$ (circles). Top: impact of quantum statistics, with green, blue and red symbols showing results for $\xi=-1$ (extrapolated), $\xi=0$, and $\xi=1$; the yellow squares show direct PIMC results for $N=14$ and $\xi=-1$. Bottom: comparing PIMC results for hydrogen to UEG results at the same parameters computed via the ESA~\cite{Dornheim_PRL_2020_ESA,Dornheim_PRB_ESA_2021}. \textcolor{red}{Not yet described ...}
%}
%\end{figure} 

\subsection{Strongly compressed beryllium\label{sec:Be_results}}

\begin{figure}\centering
\includegraphics[width=0.45\textwidth]{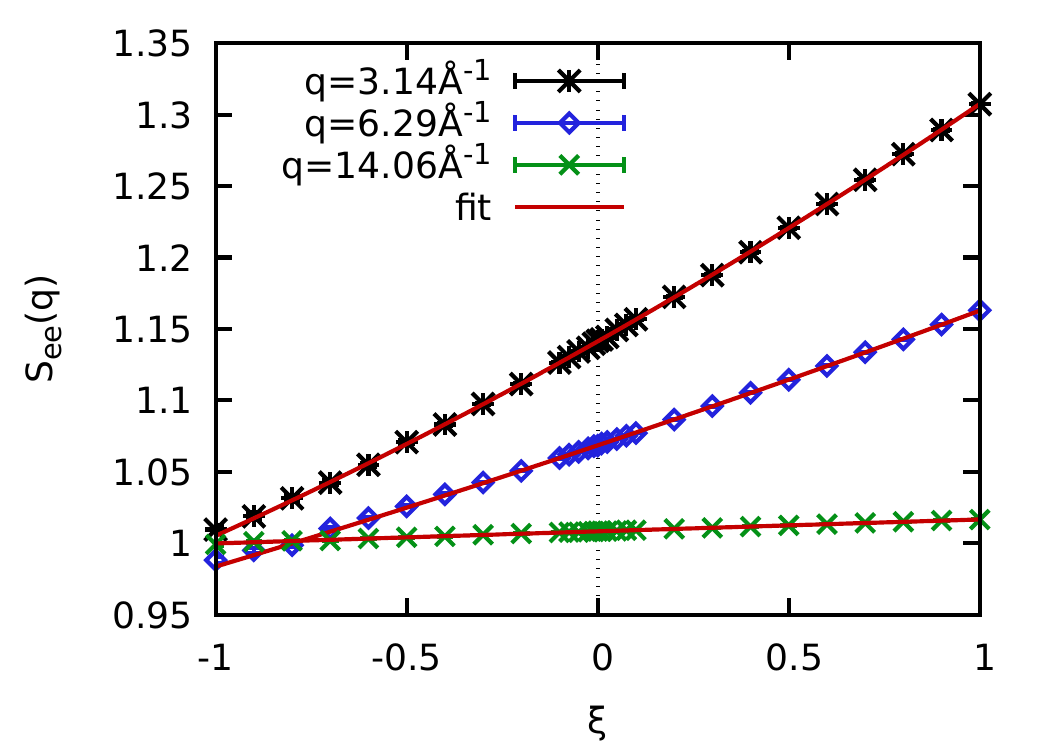}
\caption{\label{fig:Be_rs0p93_T100eV_N4_xi} The $\xi$-dependence of the electron--electron SSF $S_{ee}(\mathbf{q})$ for $N_\textnormal{atom}=4$, $r_s=0.93$, and $\Theta=1.73$. Symbols: PIMC data for $q=3.14\,$\AA${-1}$ (black stars), $q=6.29\,$\AA${-1}$ (blue diamonds), and $q=14.06\,$\AA${-1}$ (green crosses). Solid red lines: fits according to Eq.~(\ref{eq:fit}) based on PIMC data in the FSP free domain of $\xi\in[0,1]$.
}
\end{figure} 

\begin{figure}\centering
\includegraphics[width=0.425\textwidth]{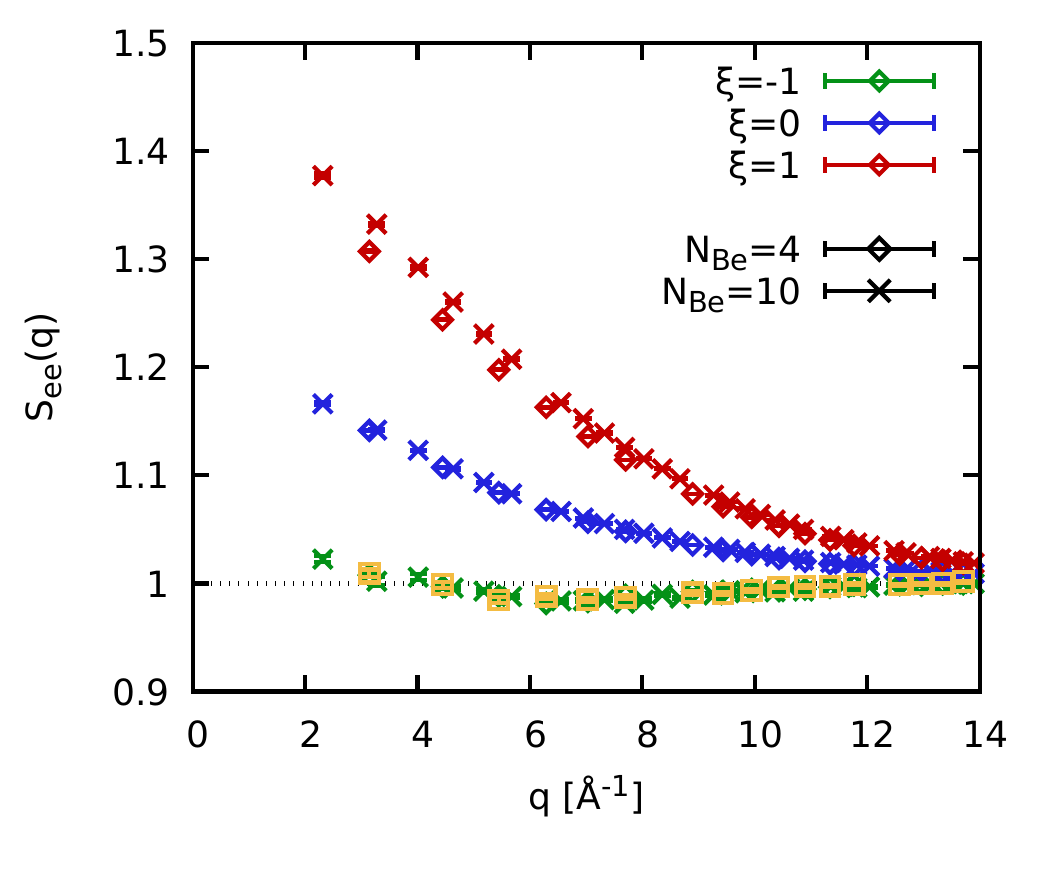}\\\vspace*{-1.25cm}
\includegraphics[width=0.425\textwidth]{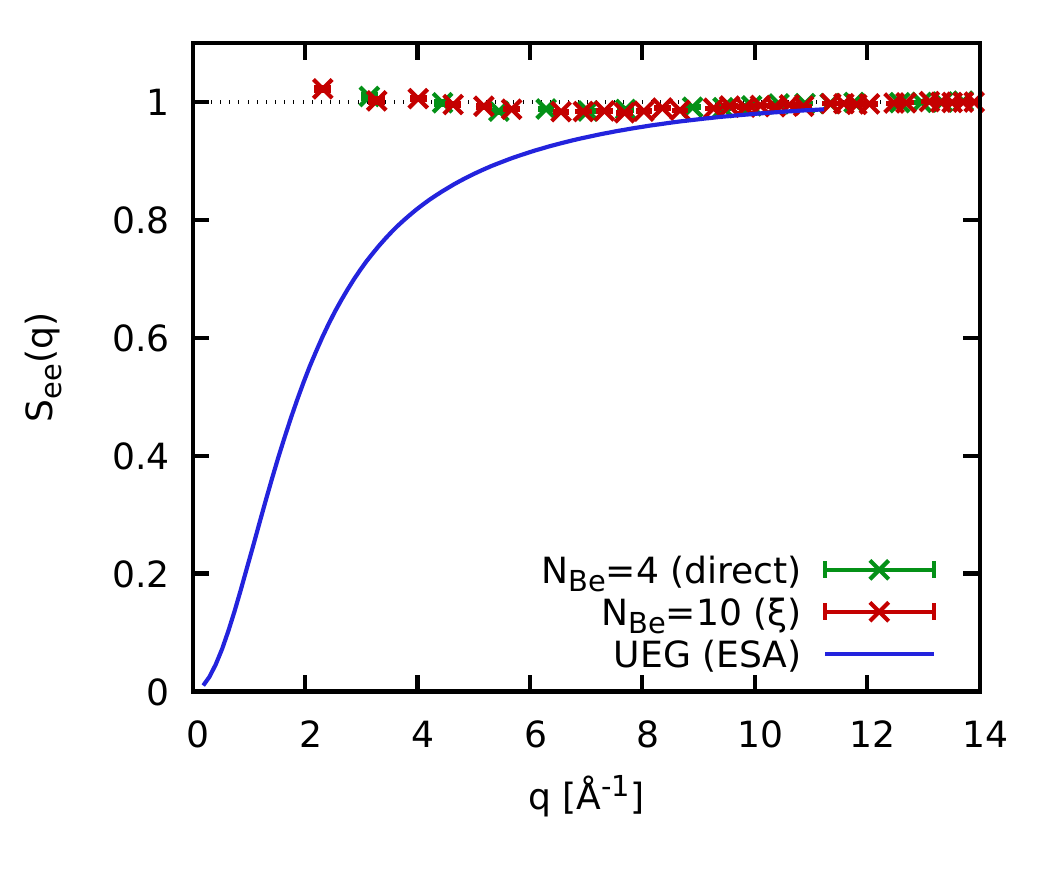}
\caption{\label{fig:Be_rs0p93_T100_N} PIMC results for the electronic SSF $S_{ee}(\mathbf{q})$ of Be $r_s=0.93$ and $\Theta=1.73$ for $N_\textnormal{atom}=4$ (diamonds) and $N_\textnormal{atom}=10$ (crosses) Be atoms. Top: impact of quantum statistics, with green, blue and red symbols showing results for $\xi=-1$ (extrapolated), $\xi=0$, and $\xi=1$; the yellow squares show direct PIMC results for $N_\textnormal{atom}=4$ and $\xi=-1$. Bottom: comparing PIMC results for hydrogen to UEG results via ESA~\cite{Dornheim_PRL_2020_ESA,Dornheim_PRB_ESA_2021}. 
}
\end{figure} 

As a final example, we consider compressed Be at $r_s=0.93$ ($\rho=7.5\,$g/cc) and $\Theta=1.73$ ($T=100\,$eV). Such conditions have been realized in experiments at the NIF~\cite{Tilo_Nature_2023,MacDonald_POP_2023}, and have recently been studied with the $\xi$-extrapolation method in Ref.~\cite{Dornheim_Science_2024}.
In Fig.~\ref{fig:Be_rs0p93_T100eV_N4}, we show an analysis of the familiar set of structural properties computed for $N_\textnormal{atom}=4$ Be atoms (i.e., $N=16$ electrons) with the usual color code. We observe the same qualitative trends as reported for hydrogen in the previous sections and restrict ourselves to a discussion of the key differences here. First and foremost, we find excellent agreement between the $\xi$-extrapolation and the exact direct PIMC simulation for $\xi=-1$ for all considered observables.
This can be discerned more directly for the electron--electron SSF in Fig.~\ref{fig:Be_rs0p93_T100eV_N4_xi}, where we explicitly show the $\xi$-dependence for three selected wavenumbers $q$.
Interestingly, $S_{ee}(\mathbf{q})$ exhibits a slightly though definite non-monotonic behaviour with a shallow minimum around intermediate $q$. It is a consequence of the interplay between the plasmon weight that decreases with $q$, and the increasing weight of the elastic feature in the long wavelength limit. 
In contrast, $F_{ee}(\mathbf{q},\beta/2)$, $S_{eI}(\mathbf{q})$, and $S_{II}(\mathbf{q})$ exhibit the same qualitative trends as reported previously for hydrogen. The spin-offdiagonal PCF $g_{ud}(r)$ shown in Fig.~\ref{fig:Be_rs0p93_T100eV_N4}e) exhibits a substantial increase towards $r\to0$; this is a direct consequence of the presence of ions with a fully occupied K-shell at these conditions, which is more pronounced in the bosonic case. This trend is completely absent for the spin-diagonal pendant $g_{uu}(r)$, where $g(r)=0$ holds by definition. At the same time, we point out the remarkable impact of quantum statistics for Be at these conditions with $g_{uu}^{\xi=1}(0)\approx4$ for bosons. We also note that we find an average sign of $S\approx0.11$ for $N_\textnormal{atom}=4$ Be atoms in our direct PIMC calculations.

Finally, we study the dependence of the electron--electron SSF on the system size in Fig.~\ref{fig:Be_rs0p93_T100_N}. Specifically, the diamonds and crosses show our PIMC results for $N_\textnormal{atom}=4$ and $N_\textnormal{atom}=10$, and the red, blue and green symbols in the top panel correspond to $\xi=1$, $\xi=0$, and $\xi=-1$. Overall, we find the same qualitative trends as observed for hydrogen in Fig.~\ref{fig:Hydrogen_rs2_theta1_N} above, namely a substantial dependence on the number of particles in the bosonic case which is absent for boltzmannons and fermions. This conclusion is further substantiated by the yellow squares that show the exact, direct fermionic PIMC simulations for $N_\textnormal{atom}=4$ and nicely agree with the $\xi$-extrapolated results for both system sizes.

\begin{figure}\centering
\includegraphics[width=0.38\textwidth]{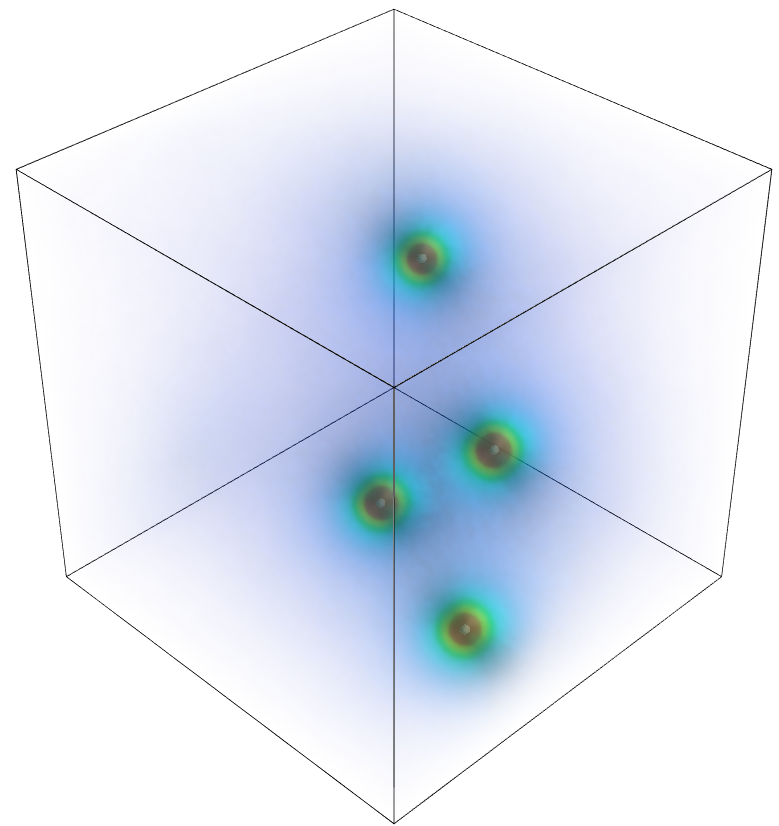}
\caption{\label{fig:3D} Electronic density distribution in a fixed ion snapshot for $N_\textnormal{atom}=4$ Be atoms at $r_s=0.93$ and $\Theta=1.73$.
}
\end{figure} 

In the bottom panel of Fig.~\ref{fig:Be_rs0p93_T100_N}, we compare our PIMC results for Be with the corresponding electronic SSF for the UEG at the same conditions; it has been computed within the aforementioned ESA~\cite{Dornheim_PRL_2020_ESA,Dornheim_PRB_ESA_2021} and is depicted by the solid blue curve. Evidently, the Be system does not even qualitatively resemble the UEG model despite the comparably high temperature. More specifically, the localized K-shell electrons, as well as the loosely localized electronic screening cloud give rise to a pronounced elastic feature in the dynamic structure factor $S_{ee}(\mathbf{q},\omega)$ in this regime~\cite{Tilo_Nature_2023,boehme2023evidence,Dornheim_Science_2024}. This, in turn, leads to an increasing normalization $S_{ee}(\mathbf{q})$ [cf.~Eq.~(\ref{eq:DSF})] for small $q$ despite the vanishing plasmon.

\begin{figure*}\centering\hspace*{-0.9cm}
\includegraphics[width=0.55\textwidth]{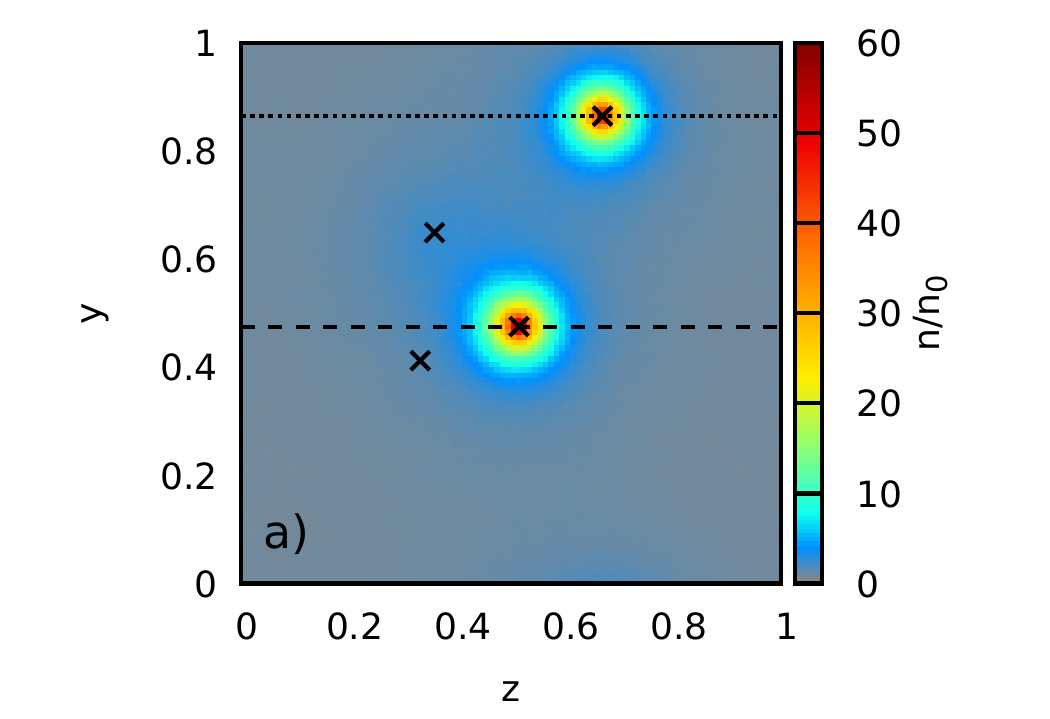}\hspace*{-0.9cm}\includegraphics[width=0.55\textwidth]{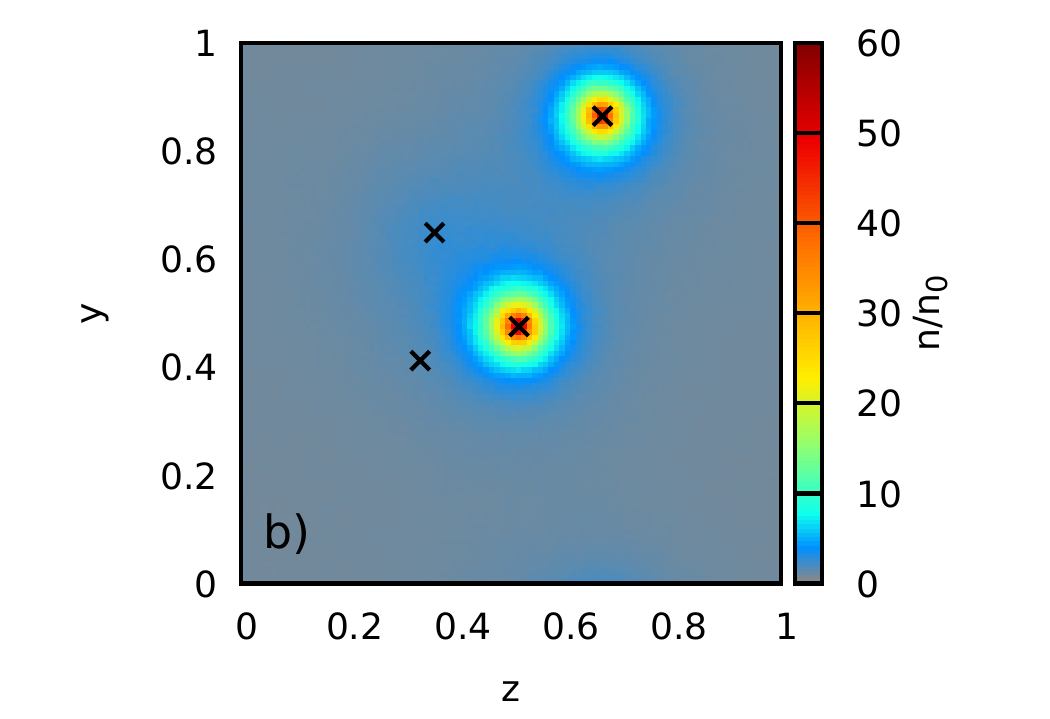}\\
\hspace*{-0.9cm}\includegraphics[width=0.55\textwidth]{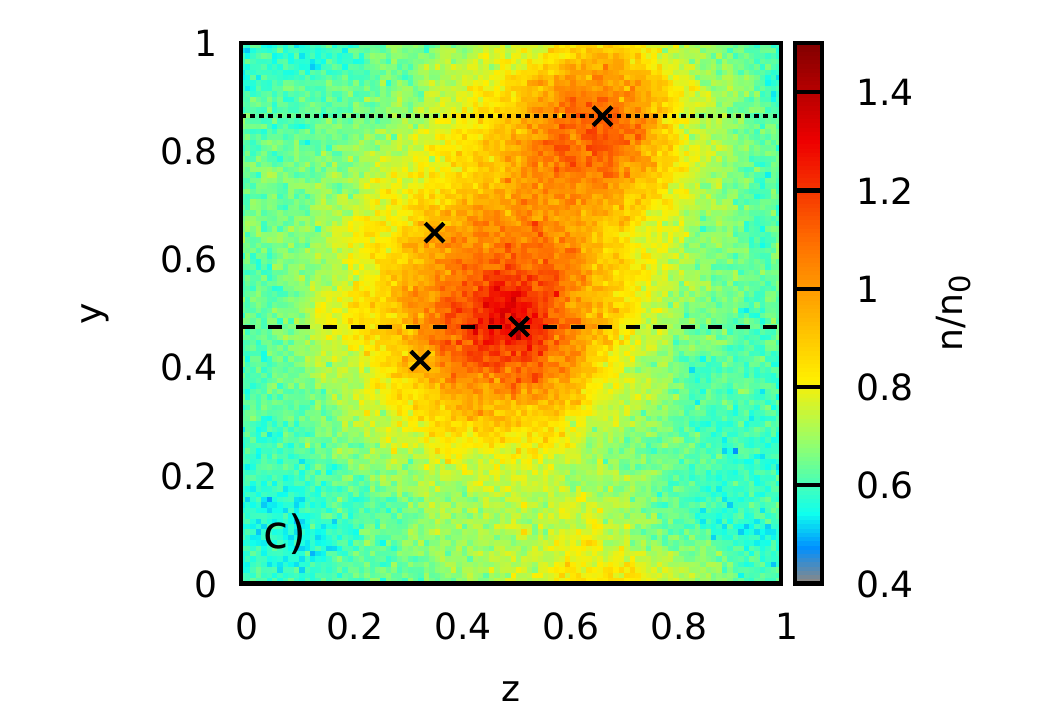}\hspace*{-0.9cm}\includegraphics[width=0.55\textwidth]{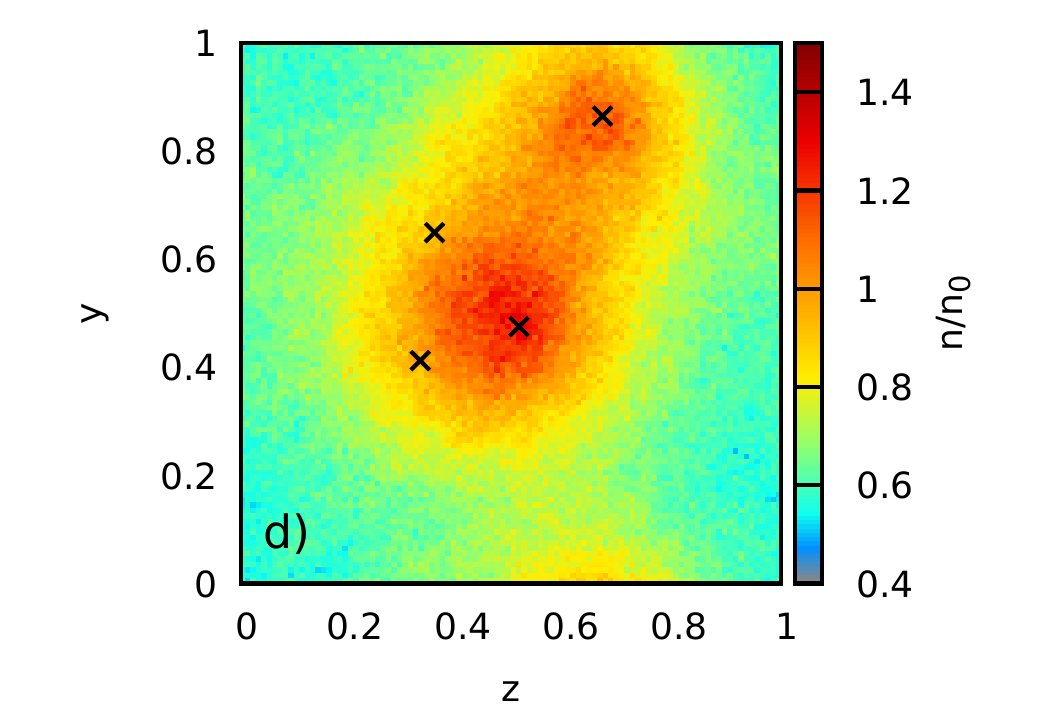}
\caption{\label{fig:Be_SNAP_Heatmap} \emph{Ab initio} PIMC results for the electronic density in the $y$-$z$-layer. Panels a) and b) have been computed for a layer with two ions in it, whereas panels c) and d) are taken from the interstitial region. The left and right columns show exact, direct fermionic PIMC calculations and corresponding $\xi$-extrapolated results based on input data from the FSP free region of $\xi\in[0,1]$, respectively. The dashed and dotted horizontal lines show scan lines that are investigated in more detail in Fig.~\ref{fig:Be_SNAP_SCAN}.
}
\end{figure*} 

\begin{figure*}\centering
\includegraphics[width=0.45\textwidth]{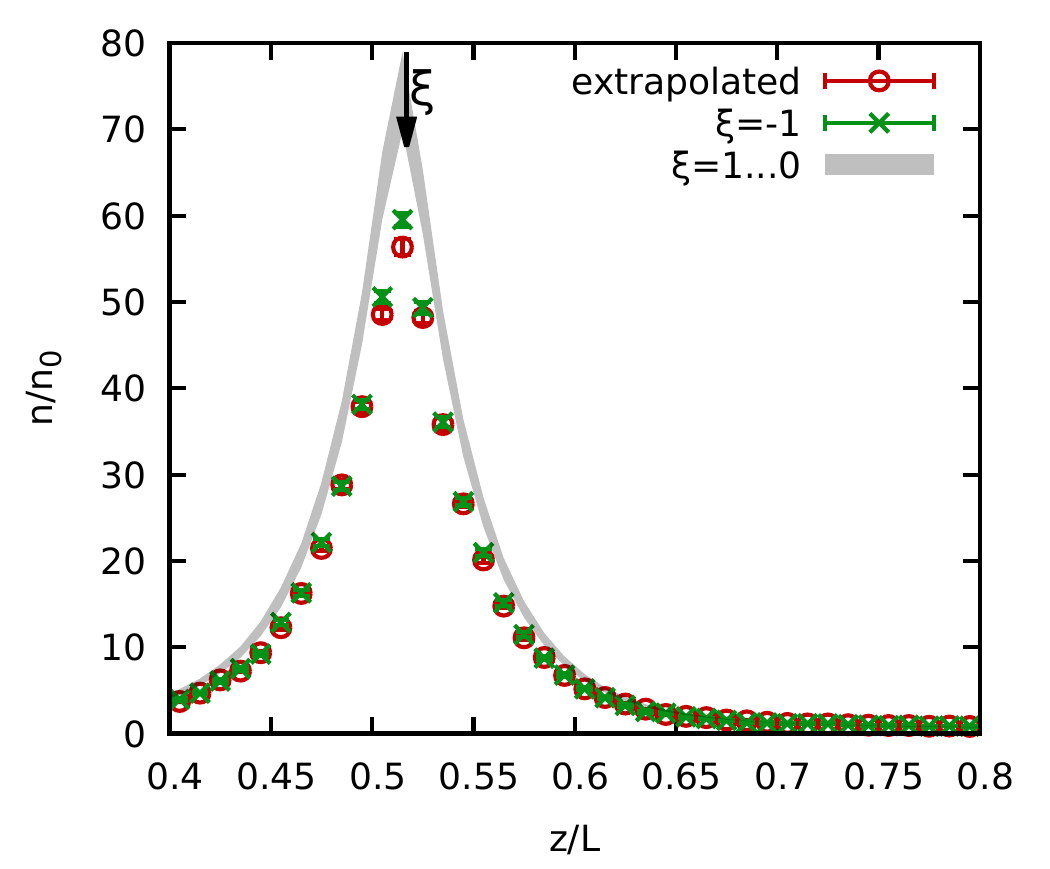}\includegraphics[width=0.45\textwidth]{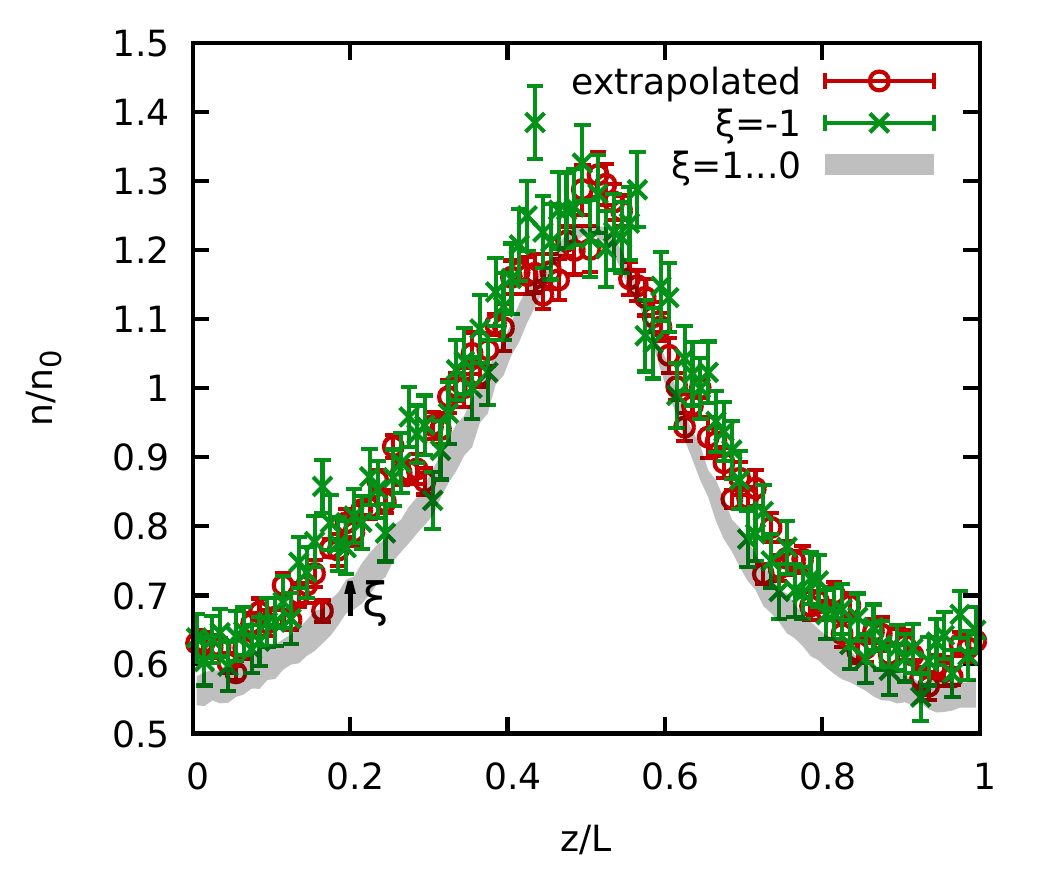}\\
\includegraphics[width=0.45\textwidth]{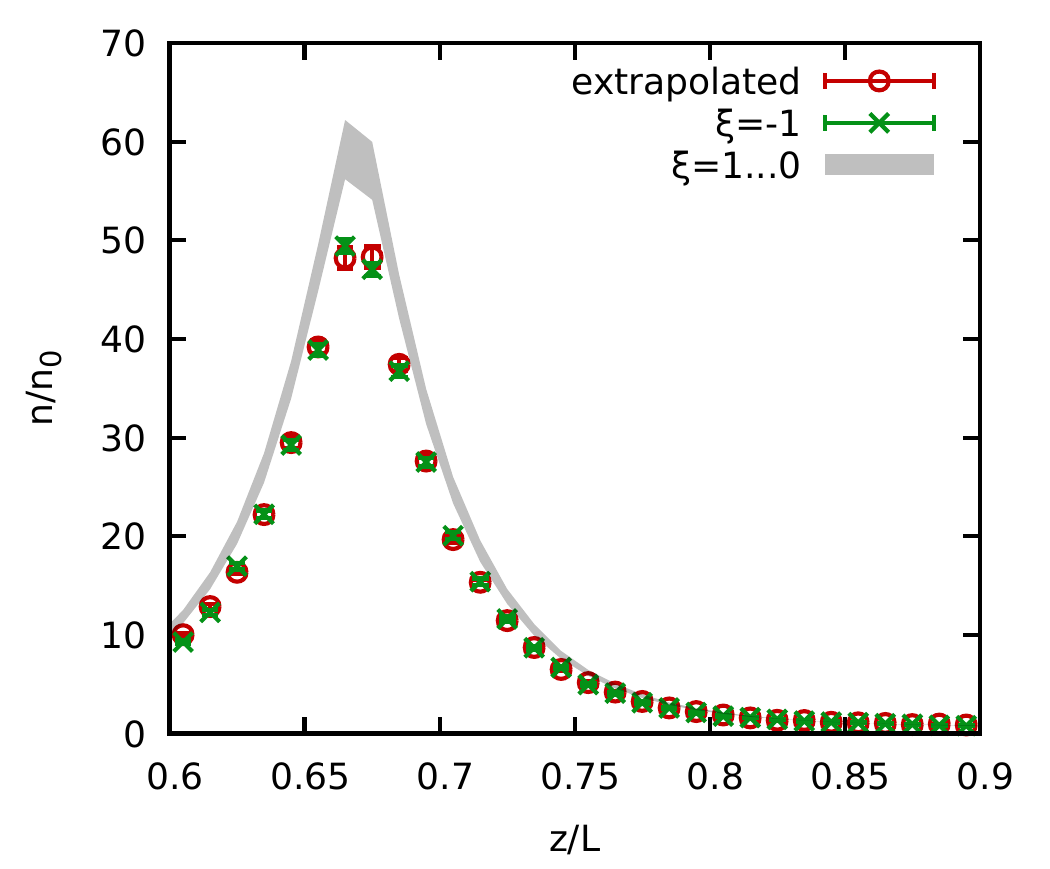}\includegraphics[width=0.45\textwidth]{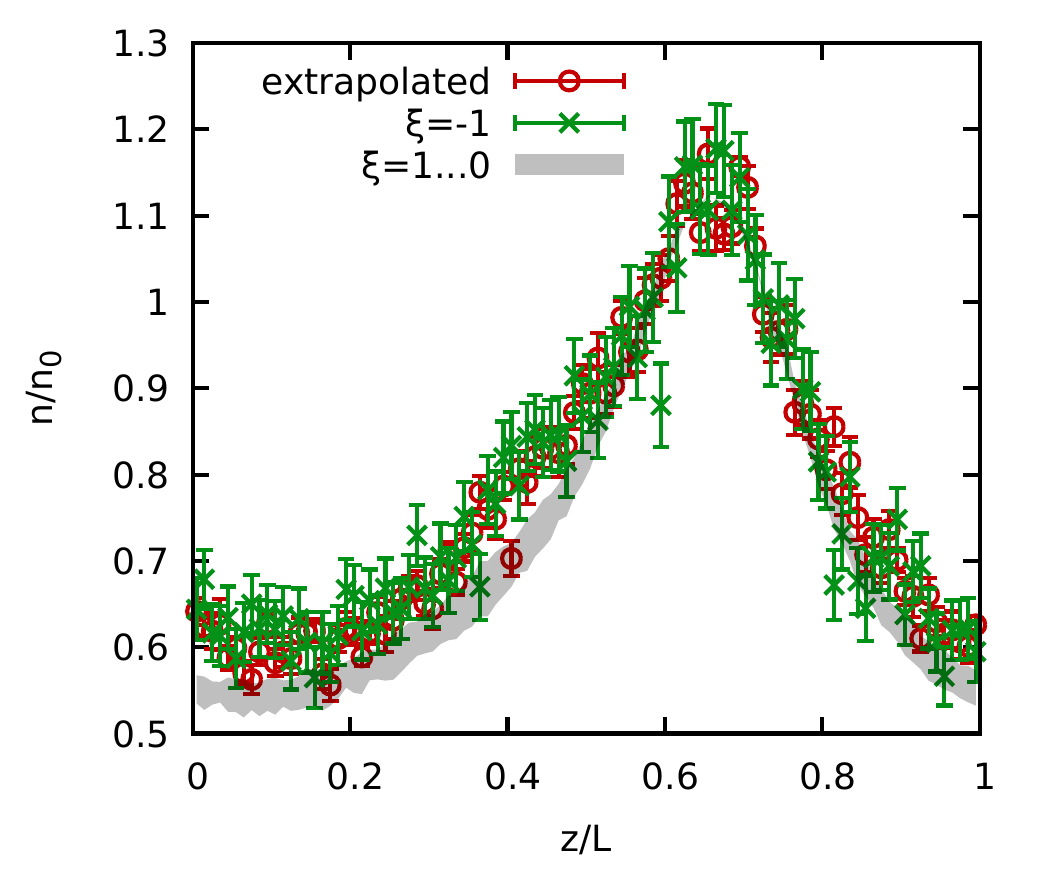}
\caption{\label{fig:Be_SNAP_SCAN} Scan lines of the electronic density for the dashed (top) and dotted (bottom) horizontal lines in Fig.~\ref{fig:Be_SNAP_Heatmap}. The left and right columns correspond to the layer with and without ions in it.}
\end{figure*}

\subsection{Strongly compressed beryllium: snapshot\label{sec:Be_SNAP_results}}

Let us conclude our investigation with a brief study of the $\xi$-extrapolation method applied to the electronic problem in a fixed external ion snapshot potential. A similar set-up has recently allowed B\"ohme \emph{et al.}~\cite{Bohme_PRL_2022,Bohme_PRE_2023} to present the first exact results for the electronic density response and the associated exchange--correlation kernel of warm dense hydrogen. In addition, such calculations provide direct information about the impact of the ions on the density response~\cite{Dornheim_PRE_2023, Moldabekov_PRR_2023, JCP_averaging}, and about the formation of bound states and molecules~\cite{Moldabekov_JCTC_BoundState_2023}.
In Fig.~\ref{fig:3D}, we show PIMC results for the electronic density computed for a snapshot of $N_\textnormal{atom}=4$ Be atoms at the same conditions studied in Sec.~\ref{sec:Be_results}.
These results nicely illustrate a high degree of electronic localization around the nuclei, and a substantially reduced density in the interstitial region. From the perspective of the $\xi$-extrapolation method that constitutes the focus of the present work, one might expect that such a spatially resolved observable constitutes a most challenging benchmark case due to the comparably larger impact of quantum statistics effects in the vicinity of the nuclei.

In Fig.~\ref{fig:Be_SNAP_Heatmap}, we show the corresponding electronic density in the $y$-$z$-plane, and the $x$-positions have been chosen so that panels a) and b) approximately contain two ions, whereas panels c) and d) show the interstitial region without such nuclei. Further, the left column shows exact, direct fermionic PIMC results, whereas the right column has been computed from the $\xi$-extrapolation via Eq.~(\ref{eq:fit}) using as input PIMC results from the sign-problem free domain of $\xi\in[0,1]$. Most importantly, we find no significant differences between the direct and the extrapolated results in both cases.
This can be seen more clearly in Fig.~\ref{fig:Be_SNAP_SCAN}, where we show scan lines over the $2D$ densities. Let us first consider the left column that has been obtained for the layer that includes two ions, and the top and bottom panels correspond to the dotted and dashed lines in Fig.~\ref{fig:Be_SNAP_Heatmap}. As it is expected, we find a comparably larger localization in the bosonic limit around the nuclei. Moreover, the $\xi$-extrapolation works very well, although there appear small differences around the maximum of the density in the top panel. This, however, is not indicative of a statistically significant systematic underestimation of the true fermionic density in this region, and does not occur for the other nuclei (i.e., see the bottom panel). The right column shows the same analysis for the layer that is located in the interstitial region. In this case, the bosonic density is comparably reduced to the other data sets; this is required to balance the increased bosonic localization around the nuclei. Most importantly, the $\xi$-extrapolation fully captures the correct fermionic limit everywhere.

\section{Summary and Discussion\label{sec:summary}}

In this work, we have presented a detailed investigation of a variety of structural properties for warm dense hydrogen and beryllium. As the first test case, we have considered hydrogen at $r_s=2$ and $\Theta=1$, where most of the electrons are assumed to be free~\cite{Moldabekov_JCTC_BoundState_2023,Filinov_PRE_2023,Militzer_PRE_2001}.
As a consequence, the system behaves qualitatively similar to the free UEG, where it is known that the $\xi$-extrapolation works well~\cite{Dornheim_JCP_2023,Dornheim_JPCL_2024}; the same does indeed hold for H at these conditions.
The second test case concerns hydrogen at solid density,  $r_s=3.23$ and $\Theta=1$. These conditions can be created in experiments with hydrogen jets~\cite{Zastrau,Fletcher_Frontiers_2022}, and might give rise to a nontrivial \emph{roton-type feature} at sufficiently high temperature~\cite{Hamann_PRR_2023,Dornheim_Nature_2022}. From a theoretical perspective, they constitute a more challenging example due to the comparably larger impact of quantum statistics that is reflected by the somewhat more severe fermion sign problem. In practice, we find that simulations based on Bose-Einstein statistics result in the formation of molecules, which are completely absent for a proper fermionic treatment of the electrons. At the same time, the $\xi$-extrapolation method nicely captures these qualitative differences and works equally well for all the investigated properties. The third physical system studied in this work concerns compressed Be at $r_s=0.93$ and $\Theta=1.73$; these conditions are relevant for experiments at the NIF~\cite{Tilo_Nature_2023,MacDonald_POP_2023} and have very recently been studied with the $\xi$-extrapolation method by Dornheim \emph{et al.}~\cite{Dornheim_Science_2024}.
Unsurprisingly, the complex interplay between the more strongly charged Be nuclei and the electrons gives rise to a richer physics that is reflected by an even more pronounced impact of quantum statistics compared to hydrogen despite the larger value of $\Theta$. This includes the partial double-occupation of the K-shell, which strongly depends on the spin-polarization of the involved electrons, giving rise to substantial differences in the spin-resolved pair correlation functions. 
Nevertheless, the $\xi$-extrapolation method works exceptionally well despite these complexities.
Finally, we have used PIMC to solve the electronic problem in the external potential of a fixed configuration of $N_\textnormal{atom}=4$ Be nuclei. From a technical perspective, analyzing the corresponding electronic density distribution might constitute the most challenging benchmark case that we have considered in this work as it allows us to spatially resolve the impact of quantum statistics, whereas potential systematic errors might be averaged out in aggregated properties such as the SSFs and PCFs considered before.
In practice, the $\xi$-extrapolation method works remarkably well both in the vicinity of the nuclei, as well as in the interstitial region.
All PIMC results that have been presented here are freely available online~\cite{repo} and can be used to benchmark new methods and approximations.

We are convinced that these findings open up a variety of potential projects for future works. While the $\xi$-extrapolation method has worked exceptionally well in all presently studied cases, previous findings for the warm dense UEG model~\cite{Dornheim_JCP_2023,Dornheim_JPCL_2024} indicate that this idea breaks down for $\Theta<1$. The particular limits will likely strongly depend both on the density and the elemental composition. A detailed study of the applicability range of the method for various light elements and their mixtures thus constitutes an important task for dedicated future works. 

A particular strength of the $\xi$-extrapolation method is given by its polynomial scaling with the system size, which combines a number of advantages. First, this access to comparably large systems allows one to study finite-size effects, which are small for the structural properties studied here, but may play a more important role for the computation of an equation-of-state~\cite{Militzer_PRE_2001,Filinov_PRE_2023}. The latter are of direct importance for a number of applications including laser fusion~\cite{hu_ICF}, and are thus of considerable interest in their own right. 
Second, simulating large systems allows one to compute properties such as the ITCF in the limit of small $q$. This is important e.g.~to describe XRTS experiments in a forward-scattering geometry~\cite{Dornheim_Science_2024}, and might be needed to fully capture dynamic long-range effects in other properties~\cite{Rygg_PRL_2023}.

An additional direction for future research is the dedicated study of density response properties. In this regard, we note the excellent performance of the $\xi$-extrapolation method w.r.t.~the ITCF $F_{ee}(\mathbf{q},\tau)$, which can be used 
as input for the imaginary-time version of the fluctuation--dissipation theorem~\cite{bowen2,Dornheim_insight_2022}
\begin{eqnarray}
    \chi_{ab}(\mathbf{q},0) = - \frac{\sqrt{N_a N_b}}{\Omega} \int_0^\beta \textnormal{d}\tau F_{ab}(\mathbf{q},\tau)\ .
\end{eqnarray}
In this way, one can get direct access to the species-resolved static density response function $\chi_{ab}(\mathbf{q},0)$ and, in this way, a set of exchange--correlation properties such as the electron--ion local field correction of different systems.
Moreover, the corresponding estimation of higher-order imaginary-time correlation functions can be used as input for similar relations~\cite{Dornheim_JCP_ITCF_2021} that give one access to a variety of nonlinear response properties~\cite{Dornheim_PRL_2020,Dornheim_PRR_2021,Dornheim_JPSJ_2021,Dornheim_CPP_2021}.

Finally, we note that the impact of quantum statistics is comparably large for Coulomb interacting systems such as in warm dense matter, but may be substantially smaller for more short-range repulsive interactions~\cite{Dornheim_NJP_2022}. This opens up the intriguing possibility to study ultracold fermionic atoms such as $^3$He~\cite{Ceperley_PRL_1992,Dornheim_SciRep_2022,Godfrin2012} with unprecedented accuracy. Given the above, returning to two-component systems, the $\xi-$extrapolation method can also improve our rather poor understanding of liquid $^3$He-$^4$He mixtures at low temperatures. Binary boson-fermion mixtures have traditionally gathered strong interest owing to the possibility of double superfluidity and the unique correlation-driven interplay between different quantum statistics~\cite{Ebner1970,Krotscheck1993,Ferrier2014}. Nevertheless, ab initio simulations of such systems are essentially non-existent. To our knowledge, finite temperature isotopic helium mixtures have been so far investigated within the fixed-node approximation~\cite{Boninsegni1995,Boninsegni1997,Moroni1998,Ujevic2023} or with the PIMC method but neglecting fermionic exchange effects~\cite{Boninsegni2018}. The quantities of interest have been limited to the superfluid fraction, the component kinetic energies, the momentum distribution functions and the pair correlation functions. Thus, the $\xi-$extrapolation technique, being sign problem free inside the phase diagram region of its applicability, can be employed for the acquisition of extensive quasi-exact thermodynamic and structural results as well as for the first reconstruction of collective excitation spectra.

\section*{Acknowledgments}
This work was partially supported by the Center for Advanced Systems Understanding (CASUS), financed by Germany’s Federal Ministry of Education and Research (BMBF) and the Saxon state government out of the State budget approved by the Saxon State Parliament. This work has received funding from the European Research Council (ERC) under the European Union’s Horizon 2022 research and innovation programme
(Grant agreement No. 101076233, "PREXTREME"). 
Views and opinions expressed are however those of the authors only and do not necessarily reflect those of the European Union or the European Research Council Executive Agency. Neither the European Union nor the granting authority can be held responsible for them. Computations were performed on a Bull Cluster at the Center for Information Services and High-Performance Computing (ZIH) at Technische Universit\"at Dresden, at the Norddeutscher Verbund f\"ur Hoch- und H\"ochstleistungsrechnen (HLRN) under grant mvp00024, and on the HoreKa supercomputer funded by the Ministry of Science, Research and the Arts Baden-W\"urttemberg and
by the Federal Ministry of Education and Research.

%%%%%%%%%%%%%%%%%%%%%%%%%%%%%%%%%%%%%%%%%%%%%%%%%%%%%%%%%%%%%%%%%%%%%%%%%%%%%%%%
% literature
%%%%%%%%%%%%%%%%%%%%%%%%%%%%%%%%%%%%%%%%%%%%%%%%%%%%%%%%%%%%%%%%%%%%%%%%%%%%%%%%
\bibliography{bibliography}

\begin{thebibliography}{158}
\bibitem [{rep()}]{repo}%
  \BibitemOpen
  \href@noop {} {}\bibinfo {note} {A link to a repository containing all PIMC raw data will be made available upon publication.}\BibitemShut {Stop}%
\end{thebibliography}
\end{document}